\documentclass[11pt]{article}
\usepackage{amssymb}
\usepackage{amsmath}
\usepackage{amscd}
\usepackage{latexsym}
\usepackage{graphics}
\usepackage{color}
\usepackage{mathtools}
\usepackage[dvipsnames]{xcolor}
\usepackage{tikz}
\usepackage[linkcolor=blue,linkbordercolor=blue,colorlinks=true, citecolor=BrickRed]{hyperref}
\usetikzlibrary{arrows}%shapes,decorations.pathreplacing}
\input xy
\xyoption{all}

\catcode`@=11
\renewcommand{\section}
{\@startsection{section}{1}{0pt}{\medskipamount}{\medskipamount}{\large\bf}}
\makeatletter\renewcommand{\subsection}
{\@startsection{subsection}{2}{\z@}{-3.25ex plus -1ex minus -.2ex}
{1.5ex plus .2ex}{\it }}

\numberwithin{equation}{section}
\catcode`@=12

\def\a{\alpha}
\def\b{\beta}
\def\g{\gamma}

\def\e{\epsilon}
\def\l{\lambda}

\def\vp{\varphi}

\def\m{\mu}
\def\n{\nu}

\def\beq{\begin{equation}}
\def\eeq{\end{equation}}
\def\bea{\begin{eqnarray}}
\def\eea{\end{eqnarray}}

\renewcommand{\e}{\,\mathrm{e}\,}
\newcommand{\id}{{\mbf 1}}
\newcommand{\im}{\,\mathrm{i}\,}
\newcommand{\diff}{\mathrm{d}}

\newcommand{\C}{{\mathbb{C}}}

\newcommand{\ca}{{\cal{A}}}
\newcommand{\cf}{{\cal{F}}}

\newcommand{\tr}{{\rm tr}}

\newcommand{\uo}{{{\rm U}(1)}}

\newcommand{\mbf}[1]{{\boldsymbol {#1} }}

\def\>{\rangle}
\def\<{\langle}
\def\+{\dagger}
\def\={\ =\ }

\newcommand{\lb}{\left(}
\newcommand{\rb}{\right)}

\newcommand{\com}[2]{\left[#1,#2\right]}

\def\ang1{$15$} % angle for arrows in quiver diagram
\def\ang2{$15$}
\def\haken{\mathbin{\hbox to 6pt{%
\vrule height0.4pt width5pt depth0pt
\kern-.4pt
\vrule height6pt width0.4pt depth0pt\hss}}}
\begin{document}

\begin{titlepage}
\setcounter{page}{0}
\begin{flushright}
ITP--UH--{12/16}
\end{flushright}

\vskip 1.8cm

\begin{center}

{\Large\bf Sasakian quiver gauge theory on the Aloff-Wallach space $X_{1,1}$}

\vspace{15mm}

{\large Jakob C. Geipel}
\\[5mm]
\noindent {\em Institut f\"ur Theoretische Physik}\\ 
{\em Leibniz Universit\"at Hannover}\\
{\em Appelstra\ss e 2, 30167 Hannover, Germany}\\
Email: {\tt jakob.geipel@itp.uni-hannover.de}

\vspace{15mm}

\begin{abstract}
\noindent
We consider the $\mathrm{SU}(3)$-equivariant dimensional reduction of gauge theories on spaces of the form $M^d \times X_{1,1}$ with $d$-dimensional
Riemannian manifold
$M^d$  and the Aloff-Wallach space  $X_{1,1}= \mathrm{SU}(3)/\mathrm{U}(1)$ 
endowed with its Sasaki-Einstein structure.  The condition of  $\mathrm{SU}(3)$-equivariance of vector bundles,
 which has already occurred in the studies of 
$\mathrm{Spin}(7)$-instantons on cones over Aloff-Wallach spaces, is interpreted in terms of quiver diagrams, and we 
construct the corresponding quiver bundles, using (parts of) the weight diagram of $\mathrm{SU}(3)$.  We consider 
three examples thereof explicitly and then compare the results with the quiver gauge theory on $Q_3 =\mathrm{SU}(3)/(\mathrm{U}(1) \times 
\mathrm{U}(1))$, the leaf space underlying the Sasaki-Einstein manifold $X_{1,1}$. Moreover, we study instanton solutions on the metric cone 
$C\lb X_{1,1}\rb$ by evaluating the Hermitian Yang-Mills equation. We briefly discuss some features of the moduli space thereof, following the main 
ideas of a treatment of Hermitian Yang-Mills instantons on cones over generic Sasaki-Einstein manifolds in the literature.

\end{abstract}

\end{center}
\end{titlepage}

{\baselineskip=11pt
\tableofcontents
}
%=========================================================
%=========================================================
%=========================================================
%=========================================================
%=========================================================
%=========================================================
\section{Introduction}

The emergence of extra dimensions in string theory and the typical ansatz for compactifications make a detailed understanding 
of higher-dimensional gauge theories desirable. Inspired by the seminal investigation of four-dimensional manifolds by self-dual 
connections \cite{AHS78}, generalized self-duality equations and instantons in higher dimensions 
have been studied \cite{Corrigan83, Ward84, Hull, noelle12}. Their significance in physics is 
evident in heterotic string theory where an instanton equation is part of the BPS equations \cite{noelle12, heterotic}.

Often the  manifolds  modelling the internal degrees of freedom are chosen as coset spaces $G/H$, and  
dimensional reduction of the gauge theory on $M^d \times G/H$ to a theory on  $M^d$ is known as 
\emph{coset space dimensional reduction} \cite{Coset92}. On those spaces one can 
demand $G$-equivariance of the vector bundles the gauge connection takes values in, and this \emph{equivariant dimensional reduction} yields
systematic restrictions  which can be depicted as \emph{quiver diagrams}, i.e. directed graphs. A detailed
mathematical treatment for K\"ahler manifolds can be found in \cite{GP02,SL2C} and  short physical reviews are  given e.g. in \cite{DS11, QGT}.

These quiver gauge theories have been studied for the K\"ahler cosets $\C P^1$ \cite{SL2C,PS06,Bismas,Dolan:2009ie}, 
$\C P^1 \times \C P^1$ \cite{LPS06}, and $\mathrm{SU}(3)/H$ \cite{Dolan:2009ie,SU3}. The odd-dimensional counterparts
 of  K\"ahler spaces are Sasaki manifolds \cite{BG}, and among them  Sasaki-Einstein manifolds \cite{Sparks} are of particular interest  
for compactifications in string theory because, by definition, their metric cones are Calabi-Yau \cite{Joyce,Greene}.
In the literature, \emph{Sasakian quiver gauge theory} has been studied on
the orbifold $S^3/\Gamma$ \cite{Lechtenfeld:2014fza}, on orbifolds $S^5/\mathbb{Z}_{q+1}$ of the five-sphere \cite{S5} 
and on the space $T^{1,1}$ \cite{T11}, the base space of the conifold. The five-dimensional Sasaki-Einstein  coset spaces as well 
as the new examples \cite{SaEin_S2S3,Martelli:2004} are of interest for  versions of the $\mathrm{AdS}/\mathrm{CFT}$ correspondence. In dimension 
seven, one can encounter the following typical  examples: the seven-sphere $S^7$, the Aloff-Wallach space $X_{1,1}$ \cite{AW75}, and also a new class of
spaces constructed in \cite{Martelli:2004}. They could play a role for compactifications of 11-dimensional supergravity. In this article we will consider 
the Sasakian quiver gauge theory on the 
Aloff-Wallach space $X_{1,1}$. The mathematical properties of the generic Aloff-Wallach spaces $X_{k,l}$ \cite{AW75} -- basically their 
$G_2$ and $\mathrm{Spin}(7)$ structure and, for the special case of $X_{1,1}$, being Sasaki-Einstein and even 3-Sasakian -- are well known \cite{CMS94, FKMS97}. 
Moreover, instanton solutions on these spaces have been constructed in \cite{AW11,Haupt15}. Due to  the special geometry, more precisely
the existence of Killing spinors, they have been intensively studied  
in M-theory or supergravity \cite{sugra}.

This article is organized as follows: Section \ref{sec:geometry} reviews the geometry of the space $X_{1,1}$, providing local coordinates, the structure equations, the 
Sasaki-Einstein properties as well as a comment on the closely related K\"ahler space $Q_3 \coloneqq \mathrm{SU}(3)/\uo \times \uo$. The subsequent section
begins with 
a short review of equivariant vector bundles over homogeneous spaces and the arising quiver diagrams. Then we study the equivariant gauge theory on $X_{1,1}$,
placing the focus on the evaluation of the equivariance condition, already known from \cite{AW11,Haupt15}, in terms of quiver diagrams. We discuss the 
general construction for the quiver diagrams associated to $X_{1,1}$ and clarify it by considering three examples with a small number of vertices. 
The resulting Yang-Mills functional of the equivariant gauge theory is provided, and the reduction to the quiver gauge theory on $Q_3$ is discussed in the last 
part of Section \ref{sec:qgt}. Subsequently, we study instanton solutions of the quiver gauge theory by evaluating the Hermitian Yang-Mills equations 
on the metric cone $C\lb X_{1,1}\rb$. We briefly sketch the techniques used by Donaldson \cite{Donaldson84} and Kronheimer \cite{Kronheimer90} for the 
discussion  of the Nahm equations and the application of those methods to Hermitian Yang-Mills instantons on generic Calabi-Yau cones \cite{MS15}. We discuss the
modifications that appear in our setup, due to using a different instanton connection in the ansatz for the gauge connection, in comparison with the general 
results of \cite{MS15}. The appendix  provides some technical details.

%=========================================================
%=========================================================
%=========================================================
%=========================================================
%=========================================================
%=========================================================

\section{Geometry of the Aloff-Wallach space $X_{1,1}$}
\label{sec:geometry}
In this section we review the geometric properties of the Aloff-Wallach space $X_{1,1}$ and its metric cone $C \lb X_{1,1}\rb$ which are necessary for the 
discussion in this article. Among the huge number of articles on the geometry of Aloff-Wallach spaces $X_{k,l}$\cite{AW75}, we follow the exposition given 
in the article \cite{AW11}, in which $G_2$ and $\mathrm{Spin}(7)$-instantons on the spaces have been considered. In particular, we employ their choice 
of $\mathrm{SU}(3)$ generators, 
structure constants and the ansatz for the gauge connections. Since we are aiming only at the Sasaki-Einstein structure of $X_{1,1}$, we will not consider general
spaces $X_{k,l}$. For details on theses structures we refer to \cite{AW11} and the references therein.

%=========================================================
%=========================================================
%=========================================================
%=========================================================
%=========================================================
%=========================================================
\subsection{Local coordinates and structure equations}

The Aloff-Wallach spaces \cite{AW75}, denoted as $X_{k,l}$, for coprime integers $k$ and $l$, are defined as quotients 
\bea
\label{eq:Xkl}
X_{k,l} \= G/H \coloneqq \mathrm{SU}\lb 3\rb / \mathrm{U} \lb 1\rb_{k,l}
\eea 
where the embedding of elements $h \in \mathrm{U} \lb 1\rb_{k,l}$ into $\mathrm{SU}(3)$ is given by
\bea
h \= \mathrm{diag} \lb \e^{\im \lb k+l\rb \vp}, \e^{-\im k \vp}, \e^{-\im l \vp}\rb.
\eea
It is known that the homogeneous space $X_{1,1}$  is not only Sasaki-Einstein but moreover admits a 3-Sasakian structure\footnote{This means that the 
(Riemannian) holonomy of the metric cone $C\lb X_{1,1}\rb$ can be reduced from $\mathrm{SU}(4)$ to $\mathrm{Sp}(2)$.}. Due to \cite{CGM} a homogeneous 
3-Sasakian manifold different from a sphere is a $\mathrm{SO}(3) \cong \mathrm{SU}(2)/\mathbb{Z}_2$ bundle over a quaternionic K\"ahler manifold; in the case 
of $X_{1,1}$ the underlying space is $\C P^2$. Using this result, we can construct local coordinates\footnote{Since we will work entirely on
Lie algebra level, a local description is sufficient for our purposes.} by starting from a local section of the fibration 
$\mathrm{SU}(3)\rightarrow \C P^2$, as it can be found e.g. in \cite{SU3,double_quiver}. Given a local patch 
 $\mathcal{U}_0 \coloneqq \left\{[w_0:w_1:w_2] \in \C P^2\ |\ w_0 \neq 0 \right\}$ of $\C P^2$, one can introduce coordinates
\bea
Y \coloneqq \begin{pmatrix}
             y_1\\
             y_2
            \end{pmatrix}
\sim \lb 
             1,  
             \frac{w_1}{w_0},  
             \frac{w_2}{w_0}\rb^{\mathrm{T}},
\eea 
and a local section of the bundle $\mathrm{SU}(3)\rightarrow \C P^2$ is given by
\bea
\label{eq:section_su3}
\C P^2 \ni Y \longmapsto V \coloneqq \frac{1}{\g}\begin{pmatrix}
               1 & \bar{Y}^{\+}\\
              -\bar{Y} & \Lambda
              \end{pmatrix} \in \mathrm{SU}(3)
\eea
with 
\bea
 \g \coloneqq \sqrt{1+\bar{Y}^{\+}\bar{Y}}, \ \  \ \Lambda \bar{Y}\=\bar{Y}, \ \  \ \bar{Y}^{\+} \Lambda \= \bar{Y}^{\+}, \ \ \ 
            \Lambda \coloneqq \g \id_2 -\frac{1}{\g+1} \bar{Y} \bar{Y}^{\+}, \ \  \ \Lambda^2 \= \g^2 \id_2 - \bar{Y}\bar{Y}^{\+}.
\eea
Furthermore, an arbitrary element $g$ of $\mathrm{SU}(2)$ can be written as
\bea
\label{eq:section_su2}
g \= \frac{1}{\lb 1+ z \bar{z}\rb^{1/2}} \begin{pmatrix}
                                          1 & -\bar{z}\\
                                          z & 1
                                         \end{pmatrix}
                                         \begin{pmatrix}
                                          \e^{\im \vp} & 0 \\
                                          0             & \e^{-\im \vp}
                                         \end{pmatrix},
\eea
where $z$ and $\bar{z}$ are stereographic coordinates on $\C P^1$. Putting both expressions (\ref{eq:section_su3}) and (\ref{eq:section_su2}) together, 
one gets a local section of the bundle $\mathrm{SU}(3) \longrightarrow X_{1,1}$ as
\small
\bea
\label{eq:local_section}
\lb y_1,y_2,z,\vp\rb \longmapsto \tilde V \coloneqq V \cdot g \= \frac{1}{\g}\begin{pmatrix}
               1 & \bar{Y}^{\+}\\
              -\bar{Y} & \Lambda
              \end{pmatrix} \frac{1}{\lb 1+ z \bar{z}\rb^{1/2}} 
\begin{pmatrix}				  1 & 0 &0\\
                                          0& 1 & -\bar{z} \\
                                          0 &z & 1 \\
                                          
                                         \end{pmatrix}
                                         \begin{pmatrix}
                                          1 & 0 &0\\
                                          0&\e^{\im \vp} & 0 \\
                                          0& 0             & \e^{-\im \vp} \\
                                      \end{pmatrix}.
\eea
\normalsize
Hence, the manifold can be locally described by the coordinates $\left\{y_1,\bar{y_1}, y_2,\bar{y_2},z, \bar{z},\vp \right\}$, and 
the Maurer-Cartan form provides $\mathrm{SU}(3)$ left-invariant \mbox{1-forms} $\Theta^{\a}$ and $e^i$, defined by
\bea
\label{eq:def_1forms}
\mathcal{A}_0 \coloneqq \tilde{V}^{-1} \diff \tilde{V} \eqqcolon \begin{pmatrix}
                                         \frac{2\im}{\sqrt{3}}e^8  & \sqrt{2}\Theta^2                    & -\sqrt{2}\Theta^{\bar{1}}\\
                                         -\sqrt{2}\Theta^{\bar{2}} & -\frac{\im}{\sqrt{3}} e^8 - \im e^7 & -\Theta^{\bar{3}}\\
                                         \sqrt{2}\Theta^1          & \Theta^3                            &-\frac{\im}{\sqrt{3}}e^8+\im e^7
                                        \end{pmatrix}.
\eea
Here we have defined the forms such that the generators of $\mathrm{SU}(3)$ (see Appendix \ref{sec:geometry_aw}) coincide with those from \cite{AW11}. Due to the flatness of the 
connection, $\diff \ca_0 + \ca_0 \wedge \ca_0=0$, one obtains the structure equations
\bea
\label{eq:structure_eqs_aw}
\nonumber \diff \Theta^1 &=& - \im e^7 \wedge \Theta^1 + \sqrt{3} \im e^8 \wedge \Theta^1 - \Theta^{\bar{2}3},\\
\nonumber \diff \Theta^2 &=& - \im e^7 \wedge \Theta^2 - \sqrt{3} \im e^8 \wedge \Theta^2 + \Theta^{\bar{1}3},\\
          \diff \Theta^3 &=&  - 2 \im e^7 \wedge \Theta^3 - 2 \Theta^{12},\\
 \nonumber \diff e^7 &=& - \im \lb \Theta^{1\bar{1}}+\Theta^{2\bar{2}}+\Theta^{3\bar{3}}\rb,\\ 
\nonumber \diff e^8 &=& \sqrt{3} \im \lb \Theta^{1\bar{1}}-\Theta^{2\bar{2}}\rb,
\eea
together with the complex conjugated equations for $\Theta^{\bar{\a}}$, $\a=1,2,3$. By construction, the group $\mathrm{U}(1)_{k,l}$ in the 
definition (\ref{eq:Xkl}) is generated by $I_8$ in (\ref{eq:generators}), and the remaining group $\mathrm{U}(1)$ inside $X_{1,1}$ is associated 
to $I_7$ and the local
coordinate $\vp$.

%=========================================================
%=========================================================
%=========================================================
%=========================================================
%=========================================================
%=========================================================
\subsection{Sasaki-Einstein structure}
\label{sec:sasaki_einstein}
Following \cite{AW11}, the Einstein metric is chosen to be
\bea
\label{eq:metric_aw}
\diff s^2_{X_{1,1}} \= g_{\m \n} e^{\m} \otimes e^{\n} \= \Theta^{1} \otimes \Theta^{\bar{1}}
             + \Theta^{2} \otimes \Theta^{\bar{2}} +\Theta^{3} \otimes \Theta^{\bar{3}} +e^7 \otimes e^7,
\eea
and the Sasaki structure is defined by declaring the forms 
$\Theta^{\a}$ to be holomorphic, $\tilde{J} \Theta^{\a} = \im \Theta^{\a}$. Here $\tilde{J}$ denotes the complex structure 
of the leaf space orthogonal to the contact direction $e^7$. Then the fundamental form $\omega$ associated to it satisfies the Sasaki condition
\bea 
2\omega \= \diff \eta \coloneqq \diff e^7 \= -{\im} \lb \Theta^{1\bar{1}}+\Theta^{2\bar{2}}+\Theta^{3\bar{3}}\rb ,
\eea
which implies that $\omega$ is the K\"ahler form of the leaf space. The metric cone $C\lb X_{1,1}\rb$ has by definition 
the metric
\bea
\label{eq:cone_metric}
\diff s_{C\lb X_{1,1}\rb}^2 \= r^2 \diff s_{X_{1,1}}^2 +\diff r \otimes \diff r \= r^2 \lb \diff s_{X_{1,1}}^2 +\frac{\diff r}{r} \otimes \frac{\diff r}{r}\rb
= r^2 \sum_{\a=1}^4 \Theta^{\a} \otimes \Theta^{\bar{\a}},
\eea
where one has defined a fourth holomorphic form
\bea
\Theta^4 \coloneqq \frac{\diff r}{r}- \im e^7.
\eea
Equation (\ref{eq:cone_metric}) establishes  the correspondence between the metric cone and the conformally equivalent cylinder\footnote{Considering the metric cone is tantamount
to studying the conformally equivalent cylinder for the discussion in this article. One can obtain an orthonormal basis by rescaling the forms 
$\tilde{e}^{\m}\coloneqq r \e^{\m}$. We will mainly use the cylinder for the description here.}.
The definition of $\Theta^4$ yields an integrable complex structure $J$ on the metric cone whose fundamental form  
\mbox{$\Omega \lb X, Y\rb \coloneqq g\lb JX, Y\rb$} is then given by
\bea
\Omega \= -\frac{\im}{2} r^2 \sum_{\a=1}^4 \Theta^{\a} \wedge \Theta^{\bar{\a}} = r^2 \omega + r \diff r \wedge e^7.
\eea
Due to the  Sasaki condition $\diff e^7 =2 \omega$ this form is closed and the cone $C\lb X_{1,1}\rb$, thus, carries a K\"ahler structure. 
For the cone to be Calabi-Yau, the holonomy $\mathrm{U}(4)$ of the K\"ahler manifold must be reduced further to $\mathrm{SU}(4)$, which is
ensured by the closure of the 4-form \cite{AW11}
\bea
\Omega^{4,0} \coloneqq r^4 \Theta^1 \wedge \Theta^2 \wedge \Theta^3 \wedge \Theta^4. 
\eea
Consequently, the geometric structure is that of a Calabi-Yau 4-fold, which implies the Sasaki-Einstein structure of $X_{1,1}$. 
As a Sasakian manifold, $X_{1,1}$ is a $\mathrm{U}(1)$-bundle over an underlying K\"ahler manifold, namely the leaf space of the foliation along the Reeb 
vector field, with fundamental form $\omega$. The K\"ahler manifold underlying $X_{1,1}$ is denoted as $Q_3$ or $\mathbb{F}_3$ \cite{SU3, double_quiver}
\beq
 \begin{tikzpicture}[scale=1]
    \node (a) at (0,0) {$X_{1,1}$};
    \node (b) at (4,0) {$Q_3 \coloneqq \frac{\mathrm{SU}(3)}{\mathrm{U}(1) \times \mathrm{U}(1)}$};
    \path[->] (a) edge node [above] {$\mathrm{U}(1)$} (b);
  \end{tikzpicture}
\eeq
From the (local) section in (\ref{eq:local_section}) one has locally $Q_3 \cong \C P^2 \times \C P^1$, and the space is described by the 
\mbox{coordinates $\left\{y_1,\bar{y_1}, y_2, \bar{y_2},z\right\}$}.

%=========================================================
%=========================================================
%=========================================================
%=========================================================
%=========================================================
%=========================================================
\section{Quiver gauge theory on $X_{1,1}$}
\label{sec:qgt}
Quiver diagrams are a powerful tool in representation theory, and this motivates their appearance in gauge theories, where the field content can be 
described by these directed graphs. In this section we will demonstrate the basic features of quiver gauge theories
by considering them on the spaces  $X_{1,1}$ and $Q_3$. We start the survey with a brief review of how quiver diagrams arise in the context of
gauge theories\footnote{Note that for us the term \emph{quiver gauge theory} always refers to the structures 
arising from the bundle equivariance. Thus our definition is not directly related to other forms of quiver gauge theories in the literature, e.g. \cite{stringy},
which are based on brane physics.} on reductive homogeneous spaces $G/H$ .

%=========================================================
%=========================================================
%=========================================================
%=========================================================
%=========================================================
%=========================================================
\subsection{Preliminaries of quiver gauge theory}

The condition generating the quiver diagrams, which we will usually refer to as \emph{equivariance condition}, can be understood from two point of views:
On the one hand, one could consider equivariant 
vector bundles in a rigorous algebraic fashion as it is done in \cite{GP02,SL2C}, purely based on the representation theory of
the Lie algebras involved. On the other hand, 
the  equivariance condition occurs quite naturally in the context of instanton studies, e.g. \cite{IP12,conical14,Bunk14,Lubbe, Popov09}, 
as invariance condition on gauge connections on reductive 
homogeneous \mbox{spaces $G/H$}.

\paragraph{Equivariant vector bundles}
We sketch the basics of equivariant vector bundles and their relation to quiver gauge theories, 
following roughly \cite{GP02,DS11}. For the application of this approach we refer also to the examples in \cite{SU3,S5}.
Let $G/H$ be a Riemannian coset space modelling the internal degrees of freedom, $M^d$ a $d$-dimensional Riemannian manifold, and let 
$\pi: \mathcal{E}\rightarrow M^d \times G/H$ be a Hermitian vector bundle\footnote{One should keep in mind 
that the fundamental objects of a gauge theory are \emph{principal bundles} $(P,p,X; K)$ with total space $P$, base space $X$, projection map $p$, 
and gauge group $K$ although we will work completely in terms of vector bundles in this article. They can be thought of as associated to the relevant 
principal bundle $P$. } of rank $k$, i.e. a vector bundle with structure group $\mathrm{U}(k)$. 
Suppose that the Lie group $G$ acts trivially on $M^d$ and in the usual way on
the coset space. Then the bundle is called \emph{$G$-equivariant} if the action of $G$ on the base space and on the total space, respectively, 
commutes with the 
projection map $\pi$ and induces isomorphisms among the fibers $\mathcal{E}_x \simeq \C ^k$.
By restriction and induction of bundles, $\mathcal{E}=G \times_H E$, $G$-equivariant bundles $\mathcal{E}\rightarrow M^d\times G/H$ 
are in one-to-one correspondence
with $H$-equivariant bundles $E\rightarrow M^d$ \cite{GP02}.

Since the action of the closed subgroup $H$ on the base space is trivial, the equivariance of the bundle implies that the fibers must carry 
representations of $H$. We assume that  these $H$-representations stem from the 
restriction of an irreducible\footnote{This assumption is not mandatory for the approach, but simplifies the situation due to the 
classification of irreducible representations of semisimple Lie algebras.} 
$G$-representation $\mathcal{D}$ which decomposes under restriction to $H$ as follows
\bea
\mathcal{D}_{|H} \= \bigoplus_{i=0}^m \rho_i
\eea 
where the $\rho_i$'s are irreducible $H$-representations. This yields an isotopical decomposition\footnote{In general, one can split 
the summands further into $E_i= \tilde{E}_i \otimes \underline{V_i}$, 
where $\underline{V_i}$ is an irreducible $H$-representation and the subgroup $H$ acts trivially on $\tilde{E}_i$ \cite{LPS06}. 
Since we consider an abelian 
subgroup $H$, the irreducible
representations are 1-dimensional, so that $H$ acts as multiple of the identity on the entire space $E_i$. For an 
example of a non-abelian subgroup $H$, consider for instance \cite{S5}.  
 } of the vector bundle $E$ as a Whitney sum 
in the very same way 
\bea
\label{eq:dec}
E \= \bigoplus_i E_i  \ \ \ \text{with} \ \ \  \lb E_i \rb_x \cong \C^{k_i} \ \ \text{carrying} \ \ \rho_i
\eea
and induces a breaking of the  generic structure group $\mathrm{U}\lb k\rb$ of the bundle to
\bea
\label{eq:breaking_structure_group}
\mathrm{U}\lb k\rb \longrightarrow \prod_{i=0}^{m} \mathrm{U} \lb k_i\rb \ \ \ \ \text{with} \ \ \ \ \sum_{i=0}^m k_i \=k.
\eea
The action of the entire group $G$ on the decomposition (\ref{eq:dec}) connects different representations $\rho_i$, i.e. it leads to homomorphisms
from $\mathrm{Hom}\lb \C^{k_i}, \C^{k_j}\rb$. In this way, the fibres of the $G$-equivariant bundle are representations of a \emph{quiver}\footnote{For 
details on representations of quiver diagrams, see for example \cite{quiver1, quiver2}}
 $\lb \mathcal{Q}_0, \mathcal{Q}_1\rb$, where $\mathcal{Q}_0$ denotes the set of vertices and $\mathcal{Q}_1$ the set of arrows.
Each vertex  $v_i \in \mathcal{Q}_0$ carries a vector space isomorphic to $\C^{k_i}$ with an $H$-representation, and the arrows are represented by linear maps 
among these spaces. The entire $G$-equivariant bundle thus carries  a representation of the quiver, and  this contruction is called a \emph{quiver bundle}.
Since the allowed arrows of the quiver diagram arise from the commutation relations of the generators with the elements of 
the subalgebra $\mathfrak{h}$, this approach is entirely based on the representation theory of $\mathfrak{h}$ and $\mathfrak{g}$, and it can be 
realized using (parts of) the weight diagram of the Lie algebra $\mathfrak{g}$.

\paragraph{Invariant gauge connections}
The equivariance condition leading to the quiver diagrams
also occurs naturally when studying instanton solutions of invariant gauge connections on reductive homogeneous spaces, 
e.g. in \cite{AW11,Lubbe,Popov09}. Let $G/H$ be a reductive homogeneous space with the $\mathrm{Ad}(H)$-invariant splitting 
\bea
\mathrm{span} \langle I_{\m}\rangle  \coloneqq \mathfrak{g} \= \mathfrak{h} \oplus \mathfrak{m} \eqqcolon \mathrm{span}\langle I_j\rangle 
\oplus \mathrm{span}\langle I_a\rangle, 
\eea
where the generators satisfy
\bea
\label{eq:comm_rel1}
\com{I_j}{I_k} \= C_{jk}^l I_l, \ \ \ \ \com{I_j}{I_a} \= C_{ja}^b I_b, \ \ \ \text{and} \ \ \ \com{I_a}{I_b}\= C_{ab}^c I_c + C_{ab}^j I_j;
\eea
the space $\mathfrak{m}$ can be identified with the tangent space of $G/H$. Let $e^{\m}$ be the 1-forms dual to the 
\mbox{generators $I_{\m}$}, which obey the structure equation
\bea
\diff e^{\m}\=-\frac{1}{2}  C_{\rho \sigma}^{\m} e^{\rho \sigma}\=-\Gamma^{\m}_{\n} \wedge e^{\n} + T^{\m},
\eea
where $\Gamma^{\m}_{\n}$ are the connection \mbox{1-forms} describing a (metric) connection $\Gamma$ on the homogeneous space, and $T^{\m}$ is its torsion.
Due to a known result from differential geometry \cite{KN} and following the approach used for example in \cite{Lubbe}, we can express a 
  $G$-invariant connection $\ca$ on the homogeneous space as
\bea
\label{eq:conn}
\ca \= I_j \otimes e^j  + X_a \otimes e^a,
\eea
where the skew-hermitian matrices $X_a$, the \emph{Higgs fields}, describe the endomorphism part. 
The connection\footnote{In principle one could also use different connections $\Gamma$ as starting point in the ansatz 
(\ref{eq:conn}). See  the comments in Section \ref{sec:metric_cone}.} $\Gamma \coloneqq I_j \otimes e^j$ takes  values entirely
in the  vertical component $\mathfrak{h}$ and is obtained by declaring the torsion to be $T(X,Y) \coloneqq - \com{X}{Y}_{\mathfrak{m}}$ for $X,Y \in 
T_e \lb G/H \rb$. 
The curvature $\cf = \diff \ca +\ca \wedge \ca$ of (\ref{eq:conn}) is then given by
\bea
\label{eq:curv}
 \cf &=& 
  \cf_{\Gamma} + \lb \com{I_j}{X_a}-C_{ja}^b X_b \rb e^j \wedge e^a + \frac{1}{2} \lb \com{X_a}{X_b}-C_{ab}^c X_c\rb e^{ab} +\diff X_a \wedge e^a.
\eea
For the connection to be $G$-invariant, terms containing the mixed \mbox{2-forms $\e^{j} \wedge e^a$} must not occur, so that one obtains -- 
assuming that the last
term in (\ref{eq:curv}) does not yield incompatible contributions\footnote{This holds true 
e.g. for constant matrices or those with $X_a = X_a \lb r\rb$, as we will consider on the 
metric \mbox{cone $C\lb G/H\rb$} in Section \ref{sec:metric_cone}.} -- the \emph{equivariance condition} \cite{Lubbe,KN}
\bea
\label{eq:eq_cond}
\com{I_j}{X_a} \= C_{ja}^b X_b.
\eea
Thus the equivariance forces the endomorphisms $X_{a}$ to act (with respect to the adjoint action) on the fibres of the bundle as the generators $I_a$
in (\ref{eq:comm_rel1}) do.

\paragraph{Construction procedure}
 Based on the outline above, we can construct an equivariant gauge connection and the corresponding quiver bundle for 
$X_{1,1}=\mathrm{SU}(3)/\mathrm{U}(1)_{1,1}$ in the following way. Let 
\bea
\label{eq:vect_dec}
\C^k \= \lb \C^{k_0}, \C^{k_1}, \ldots, \C^{k_m}\rb^{\mathrm{T}}
\eea 
be 
a decomposition of the representations on the fibres 
in $(m+1)$ terms, which yields the breaking of the structure group (\ref{eq:breaking_structure_group}) and the isotopical decomposition as
in (\ref{eq:dec}). Since the irreducible representations $\rho_i$ of the abelian subgroup $H= \mathrm{U}(1)_{1,1}$ are 1-dimensional, the group $H$ acts as
\bea
\lb \zeta_0 \ \id_{k_0}, \zeta_1\ \id_{k_1}, \ldots, \zeta_m \ \id_{k_m}\rb
\eea
on the vectors (\ref{eq:vect_dec}). The constants $\zeta_i$ can be obtained from an irreducible representation of the $\mathrm{U}(1)_{1,1}$-generator
on an $(m+1)$-dimensional vector space. This fact and the way how the quiver diagrams arise motivate to consider the gauge connection as a block matrix
of size $(m+1)^2$, whose structure is determined by the $(m+1)$-dimensional $G$-representation in which the entries are (implicity) replaced by 
endomorphisms. By construction and due to the equivariance condition (\ref{eq:eq_cond}), the quiver diagram is then  based on 
(parts of) the underlying weight diagram of the chosen $G$-representation. If the subgroup $H$ is
a maximal torus, the quiver coincides with the weight diagram because all Cartan generators occur as operators $I_j$ in (\ref{eq:eq_cond}).
 For smaller subgroups there might be 
degeneracies as double arrows in the diagram, while larger groups require a collapsing of vertices in the weight diagram along the action of the 
ladder operators of $\mathfrak{h}$ as it is done, for instance, in \cite{QGT,SU3}. 
 We will clarify this procedure for the abelian subgroup $H=\mathrm{U}(1)_{1,1}$ in the following.

%=========================================================
%=========================================================
%=========================================================
%=========================================================
%=========================================================
%=========================================================

\subsection{Equivariance condition and quiver diagrams of $X_{1,1}$}
The aforementioned approach is now applied to the space $X_{1,1}$. Following the outline above and
according to  (\ref{eq:conn}), we write 
an $\mathrm{SU}(3)$-invariant connection $\ca$ on $M^d \times X_{1,1}$ as \cite{AW11}
\bea
\label{eq:inv_conn}
 \ca \= A + I_8 \otimes e^8 + \sum_{a=1}^7 X_{a} \otimes e^a \eqqcolon A +I_8 \otimes e^8 + \sum_{\a=1}^3 \lb Y_{\a} \otimes \Theta^{\a} 
+ Y_{\bar{\a}} \otimes\Theta^{\bar{\a}}\rb  +X_7 \otimes e^7,
\eea
where $A$ is a connection on $M^d$. Moreover,  we have defined complex endomorphisms 
\bea
Y_1 \coloneqq \frac{1}{2} \lb X_1 +\im X_2\rb, \ \ \ \ Y_2 &\coloneqq& \frac{1}{2} \lb X_3 +\im X_4\rb, 
\ \ \ \ Y_3 \coloneqq \frac{1}{2} \lb X_5 +\im X_6\rb\\
\nonumber \text{with} \ \ \ \ \ Y_{\bar{\a}} &\coloneqq& - Y_{\a}^{\+}. 
\eea

In terms of the structure constants (\ref{eq:struc_const_aw}) the field strength of the connection $\ca$ is given by \cite{AW11}
\bea
\label{eq:fieldstrength}
\nonumber      
\cf &=& \diff A + A \wedge A + \lb \diff Y_{\a} + \com{A}{Y_{\a}}\rb\wedge \Theta^{\a}  
                  +\lb \diff Y_{\bar{\a}}+\com{A}{Y_{\bar{\a}}} \rb\wedge \Theta^{\bar{\a}} + 
                   \lb \diff X_7 +\com{A}{X_7}\rb \wedge e^7\\
\nonumber      &&+ \frac{1}{2} \lb \com{Y_{\a}}{Y_{\b}}- C_{\a\b}^{\g} Y_{\g}\rb \Theta^{\a\b}
             + \lb \com{Y_{\a}}{Y_{\bar{\b}}} -C_{\a \bar{\b}}^{\g} Y_{\g} -C_{\a \bar{\b}}^{\bar{\g}} Y_{\bar{\g}} + \im C_{\a\bar{\b}}^7 X_7 
               + \im C_{\a\bar{\b}}^8 I_8\rb \Theta^{\a \bar{\b}}\\
\nonumber      &&+ \frac{1}{2}\lb \com{Y_{\bar{\a}}}{Y_{\bar{\b}}}- C_{\bar{\a}\bar{\b}}^{\bar{\g}} Y_{\bar{\g}}\rb \Theta^{\bar{\a} \bar{\b}}
            + \lb \com{X_7}{Y_{\a}}-\im C_{7\a}^{\b}Y_{\b}\rb e^7 \wedge \Theta^{\a}
                + \lb \com{X_7}{Y_{\bar{\a}}}- \im C_{7\bar{\a}}^{\bar{\b}}Y_{\bar{\b}}\rb e^7 \wedge \Theta^{\bar{\a}}\\
    &&+ \lb \com{I_8}{Y_{\a}}-\im C_{8\a}^{\b} Y_{\b}\rb e^8 \wedge \Theta^{\a}  
                 +\lb \com{I_8}{Y_{\bar{\a}}}-\im C_{8\bar{\a}}^{\bar{\b}} Y_{\bar{\b}}\rb e^8 \wedge \Theta^{\bar{\a}} +  \com{I_8}{X_7}e^{87}. 
\eea
Following some notation in the literature, e.g. in \cite{SU3}, we call
\bea
\phi^{(\a)} \coloneqq Y_{\bar{\a}} \ \ \text{for} \ \ \a=1,2,3, \ \ \ \text{and} \ \ \ \ X_7
\eea
the \emph{Higgs fields} and set\footnote{As mentioned above, we implicity interpret the numbers in the Cartan generators as numbers times identity operators.}
\bea
\hat{I}_8 \coloneqq -\sqrt{3} \im I_8 \= \mathrm{diag}\lb 2,-1,-1\rb \  \ \ \ \  \ \text{and} \ \ \ \ \
\hat{I}_7 \coloneqq -\im I_7 \= \mathrm{diag} \lb 0,-1,1\rb.
\eea
The equivariance condition (\ref{eq:eq_cond}), equivalent to the vanishing of the terms in the last line of (\ref{eq:fieldstrength}), then reads
 \bea
\label{eq:equivariance_aw}
 \com{\hat{I}_8}{\phi^{(1)}} &=& 3 \phi^{(1)}, \ \ \ \ \ 
     \com{\hat{I}_8}{\phi^{(2)}}\= -3\phi^{(2)}, \ \ \text{and} \ \
     \com{\hat{I}_8}{\phi^{(3)}} \= 0 \= \com{\hat{I}_8}{X_7}.
\eea
Consequently, the endomorphisms $\phi^{(1)}$ and $\phi^{(2)\+}$ will have the same block form
and  the form of $\phi^{(3)}$ coincides with that of $X_7$, but their entries are still arbitrary and not related to each other.
The commutation relations (\ref{eq:equivariance_aw}) provide the action of the Higgs fields on the quantum numbers $(\n_7,\n_8)$ associated to 
the two Cartan generators $\hat{I}_7$ and $\hat{I}_8$ of $\mathrm{SU}(3)$
\bea
\label{eq:eq_rules_aw}
\nonumber \phi^{(1)}: \lb \n_7,\n_8\rb &\longmapsto&(\ast,\n_8+3),\\
          \phi^{(2)}: \lb \n_7,\n_8 \rb &\longmapsto& (\ast,\n_8-3),\\
\nonumber \phi^{(3)}: \lb \n_7,\n_8\rb &\longmapsto& (\ast,\n_8),\\
\nonumber  X_7:       \lb \n_7,\n_8 \rb &\longmapsto& (\ast,\n_8).
\eea 
Since the quantum number $\n_7$ does not enter the equivariance condition, it is reasonable\footnote{This corresponds to the 
isotopical decomposition (\ref{eq:dec}): The representation $\mathcal{D}$ is restricted under the subgroup  $\mathrm{U}(1)_{1,1}$
rather than under a maximal torus $\mathrm{U}(1) \times \mathrm{U}(1)$.} to label the vertices in the quiver diagram only by
the number $\n_8$, so that one obtains effectively a modified version of the holomorphic chain \cite{SL2C}: a diagram consisting of double arrows between 
adjacent vertices and double loops at each vertex,
\beq
    \begin{tikzpicture}[->,scale=2.7]
    \node (a) at (0,0) {$\bullet$};
    \node (la) at (0,-.17){$\mathbf{(p-3{m})}$};
    \node (b) at (1,0) {$\bullet$};
    \node (lb) at (1,-.17) {$\mathbf{(p-3{m}+3)}$};
    \node (c) at (2,0) {$\ldots$};
    \node (d) at (3,0) {$\bullet$};
    \node (ld) at (3,-.17) {$\mathbf{(p-3)}$};
    \node (e) at (4,0) {$\bullet$};
    \node (le) at (4,-.17) {$\mathbf{(p)}$};

    \path[ultra thick,->>](a.east) edge node {} (b.west);
    \path[ultra thick,->>](b.east) edge node {} (c.west);
    \path[ultra thick,->>](c.east) edge node {} (d.west);
    \path[ultra thick,->>](d.east) edge node {} (e.west);
  
    \draw [ultra thick,->>,blue](a) to [out=60,in=120,looseness=13] (a);
    \draw [ultra thick,->>,blue](b) to [out=60,in=120,looseness=13] (b);
    \draw [ultra thick,->>,blue](d) to [out=60,in=120,looseness=13] (d);
    \draw [ultra thick,->>,blue](e) to [out=60,in=120,looseness=13] (e);
 \end{tikzpicture}
\eeq
where the  black two headed arrows denote the contributions by $\phi^{(1)}$ and $\phi^{(2)\+}$, while the endomorphisms 
$\phi^{(3)}$ and $X_7$ are represented by the blue two headed loops\footnote{Using  one arrow with two heads as symbol for two arrows improves the readibility
of the more complicated diagrams like Figure \ref{fig:adj} significantly.}. Here, the integer $p$ denotes the highest weight (with respect to $ \nu_8$) of the 
representation $\mathcal{D}$. The endomorphism part of the invariant connection associated to this modified 
holomorphic chain of length ${m}+1$ is then given by
\bea
\label{eq:eff_conn1}
X_a e^a= \begin{pmatrix}
          \Psi_{p}  & \Phi_{p-3} & 0 & \ldots & 0\\
         -\Phi_{p-3}^{\+} &  \Psi_{p-3} &  \Phi_{p-6} & \ldots & \vdots\\
         0            &  -\Phi_{p-6}^{\+} & \ddots & \ddots & 0\\
         \vdots   &          \vdots        & \ddots & \Psi_{p-3}& \Phi_{p-3{m}}\\
         0  & \ldots &       0 & -\Phi_{p-3{m}}^{\+} &\Psi_{p-3{m}} 
        \end{pmatrix},
\eea
\normalsize
where we have defined the abbreviations
\bea
\Phi_{p-3j} \coloneqq \phi^{(1)}_{p-3j} \otimes \Theta^{\bar{1}}- \phi^{(2)\ \+}_{p-3j}\otimes \Theta^{{2}}
\ \ \ \ \ \text{and} \ \ \ \ \ \Psi_{p-3j}\coloneqq \phi^{(3)}_{p-3j} \otimes \Theta^{\bar{3}} + \lb X_7\rb_{p-3j} \otimes e^7 
\eea 
for $j=0,1, \ldots, {m}$, and the indices label the tail of the arrow.
The remaining contribution to the invariant connection (\ref{eq:inv_conn}) is given by the diagonal parts
\bea
\label{eq:eff_conn2}
A +\Gamma \= \mathrm{diag}\lb  \id_{k_l} \otimes  \frac{3l-p}{\sqrt{3}} \im e^8  + A_{p-3l}\rb_{l=0, \ldots,{m}},
\eea
where $A_{p-3l}$ is a component -- according to the isotopical decomposition (\ref{eq:dec}) of the fibres  -- of a connection on the bundle $E \rightarrow M^d$. 
Equations (\ref{eq:eff_conn1}) and (\ref{eq:eff_conn2}) describe the  general solution. For comparisons with gauge theories of similar 
geometric structures like $Q_3$,
it is advantageous to consider not only the decomposition under the subgroup $H$, i.e. labelling the 
vertices only by $\n_8$ as we did, but to study the equivariance conditions in the entire weight diagram of $G$. Since the weight 
diagrams\footnote{For representation theory of $\mathfrak{su}(3)$ see e.g. \cite{FH}.}  of the relevant Lie algebras are well-known, 
one can quickly construct the invariant connection by implementing the rules (\ref{eq:eq_rules_aw})
and can then project to the relevant quantum numbers.
In the following, we will consider the triangular/hexagonal weight diagram 
 of $\mathrm{SU}(3)$, 
spanned by the root system
\beq
    \begin{tikzpicture}[->,scale=1.]
    
    \node (o) at (0,0){};
    \node (a) at (-1.5,2.1) {$(-1,3)$};
    \node (b) at (-1.5,-2.1) {$(-1,-3)$};
    \node (c) at (-3,0) {$(-2,0)$};

   \path[ultra thick] (o) edge node [above right] {$I_{\bar{1}}$} (a);
   \path[ultra thick] (o) edge node [above left] {$I_{\bar{2}}$} (b);
   \path[ultra thick] (o) edge node [above] {$I_{\bar{3}}$} (c);

\end{tikzpicture}
\eeq

%=========================================================
%=========================================================
%=========================================================
%=========================================================
%=========================================================
%=========================================================

\subsubsection{Examples}
\begin{figure}[t!]
\begin{center}
    \begin{tikzpicture}[->,scale=2.7]
    \node (a) at (0,0) {$\mathbf{(0,2)}$};
    \node (b) at (.5,.866) {$\mathbf{(1,-1)}$};
    \node (c) at (-.5,.866) {$\mathbf{(-1,-1)}$};

    \path[ultra thick,->>](b) edge node {} (a);
    \path[ultra thick,->>](c) edge node {} (a);
     \path[<<->>,ultra thick, blue](c) edge node {} (b);
 
      \draw [ultra thick,->>,blue](b) to [out=60,in=120,looseness=7] (b);
      \draw [ultra thick,->>,blue](c) to [out=60,in=120,looseness=7] (c);
      \draw [ultra thick,->>,blue](a) to [out=-120, in=-60,looseness=7] (a);

     \node (d) at (2,0) {$\mathbf{(2)}$}; 
     \node (e) at (2.5,.866) {$\mathbf{(-1)}$};
     \path[ultra thick,->>](e) edge node {} (d);
     \draw [ultra thick,->>,blue](e) to [out=60,in=120,looseness=7] (e);
      \draw [ultra thick,->>,blue](d) to [out=-120, in=-60,looseness=7] (d);
     
      \node (f) at (.7,.866) {};
      \node (g) at (2.3,.866) {};
      \node (h) at (.2,0) {};
      \node (i) at (1.8,0) {};
      
      \path[dotted, thick](f) edge  node [below]{projection} (g);
      \path[dotted, thick](h) edge  node [below]{projection} (i);
      
\end{tikzpicture}
\end{center}
\caption{Quiver diagram of $X_{1,1}$ for the fundamental representation $\underline{\mathbf{3}}$ of $\mathrm{SU}(3)$: The left diagram stems from the 
implementation 
of the equivariance condition in the weight diagram of
$\mathrm{SU}(3)$  and the right one is the holomorphic chain with the loop modification and the double arrows, 
obtained from the projection by forgetting about the 
second quantum number $\n_7$, i.e. identifying points along horizontal lines.}
\label{fig:fund}
\end{figure}
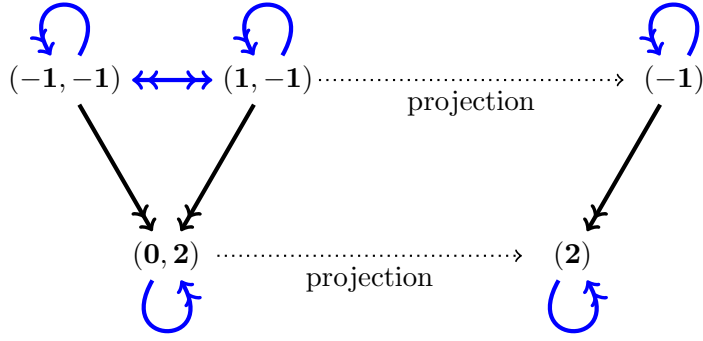

We consider three explicit examples of $\mathrm{SU}(3)$ representations and the quiver diagrams associated to them.
% \begin{itemize}\item {\bf Fundamental representation}:
\paragraph{Fundamental representation}  
Applying the prescription (\ref{eq:eq_rules_aw}) to the single triangle  of
the weight diagram of the defining representation $\underline{\mathbf{3}}$ provides the quiver diagram in \mbox{Figure \ref{fig:fund}}.
Of course, this diagram could be also obtained by direct evalution of the commutator of $\hat{I}_8=\mathrm{diag}(2,-1,-1)$ 
with an arbitrary $3\times 3$-matrix $\lb \bullet \rb$
\bea
\small
\com{\hat{I}_8}{\lb \bullet \rb} \= \begin{pmatrix}
                                                       \phantom{-} 0 \bullet & 3 \bullet & 3\bullet \\
                                                       -3 \bullet    & 0\bullet & 0\bullet\\
                                                        -3 \bullet    & 0\bullet & 0\bullet\\
                                                      \end{pmatrix}.
\eea  
\normalsize
Then the equivariance condition requires the Higgs fields to be of the form
\small
\bea
\phi^{(1)} \= \begin{pmatrix}
            0 & \ast & \ast \\
            0 & 0 & 0\\
            0 & 0 & 0
           \end{pmatrix}, \ \  
\phi^{(2)} \=\begin{pmatrix}
            0 & 0 & 0\\
           \ast & 0 & 0\\
           \ast & 0 & 0
           \end{pmatrix}, \ \  
\phi^{(3)} \= \begin{pmatrix}
            \ast & 0 & 0 \\
            0 & \ast & \ast\\
            0 & \ast & \ast
           \end{pmatrix}, \ \ 
X_7 \= \begin{pmatrix}
            \ast & 0 & 0 \\
            0 & \ast & \ast\\
            0 & \ast & \ast
           \end{pmatrix},
\eea
\normalsize
which again yields the quiver diagram Figure \ref{fig:fund}. This translates into the invariant gauge connection
\small
\bea
\ca_{\underline{\mathbf{3}}} = \begin{pmatrix}
          \frac{2}{\sqrt{3}} \im e^8 \otimes \id + \Psi_{0,2;0,2}& \Phi_{-1,-1;0,2}  &  \Phi_{1,-1;0,2}\\
          -\Phi_{-1,-1;0,2}^{\+} & -\frac{1}{\sqrt{3}} \im e^8 \otimes \id  +\Psi_{-1,-1;-1,-1}& +\Psi_{1,-1;-1,-1} \\
           -\Phi_{1,-1;0,2}^{\+} &-\Psi_{1,-1;-1,-1}^{\+} & -\frac{1}{\sqrt{3}} \im e^8 \otimes \id  +\Psi_{1,-1;1,-1}
        \end{pmatrix},
\eea
\normalsize
where we have defined 
\bea
\nonumber &&\Phi_{i,j;k,l} \coloneqq (\phi^{(1)})_{i,j;k,l} \otimes \Theta^{\bar{1}} - (\phi^{(2)\+})_{i,j;k,l} \otimes \Theta^{{2}},\\
&& \Psi_{i,j;k,l} \coloneqq (\phi^{(3)})_{i,j;k,l} \otimes \Theta^{\bar{3}} + (X_7)_{i,j;k,l} \otimes e^7;
\eea
the $\mathrm{U}(1)\times \mathrm{U}(1)$-charges $(i,j)$ denote the tail of the arrow, and $(k,l)$ its head. Going back to the 
effective quiver diagram, i.e. the modified holomorphic chain,  yields
\bea
{\ca}_{\mathbf{3}} &=& \begin{pmatrix}
          \frac{2}{\sqrt{3}} \im e^8 \otimes \id + \Psi_{2;2}&  \tilde{\Phi}_{-1;2}\\
          -\tilde{\Phi}_{-1;2}^{\+} & -\frac{1}{\sqrt{3}} \im e^8 \otimes \id  +\tilde{\Psi}_{-1;-1}\\
        \end{pmatrix},
\eea
which agrees with the general result (\ref{eq:eff_conn1}). The anti-fundamental representation $\bar{\underline{\mathbf{3}}}$, of course, leads to an
analogous diagram and connection.

\begin{figure}[t!]
\begin{center}
\small
    \begin{tikzpicture}[->,scale=2.7]
    \node (a) at (0,0) {$\mathbf{(0,4)}$};
    \node (b) at (.5,.866) {$\mathbf{(1,1)}$};
    \node (c) at (-.5,.866) {$\mathbf{(-1,1)}$};
    \node (d) at (1,1.732) {$\mathbf{(2,-2)}$};
    \node (e) at (0,1.732) {$\mathbf{(0,-2)}$};
    \node (f) at (-1,1.732) {$\mathbf{(-2,-2)}$};

     \path[ultra thick,<<-](a) edge node {} (b);
     \path[ultra thick,<<-](a) edge node {} (c);
     \path[ultra thick,<<-](b) edge node {} (d);
     \path[ultra thick,<<-](b) edge node {} (e);
     \path[ultra thick,<<-](b) edge node {} (f);
     \path[ultra thick,<<-](c) edge node {} (d);
     \path[ultra thick,<<-](c) edge node {} (e);
     \path[ultra thick,<<-](c) edge node {} (f);

     \path[<<->>,ultra thick,  blue](c) edge node {} (b);
     \path[<<->>,ultra thick,  blue](f) edge node {} (e);
     \path[<<->>,ultra thick,  blue](e) edge node {} (d);
     \path[<<->>,ultra thick,  blue](f) edge [bend left=-25] node {} (d);
 
     \draw [ultra thick,->>,blue](d) to [out=60,in=120,looseness=7] (d);
     \draw [ultra thick,->>, ,blue](e) to [out=60,in=120,looseness=7] (e);
     \draw [ultra thick,->>, blue](f) to [out=60,in=120,looseness=7] (f);
      \draw [ultra thick,->>, blue](a) to [out=-120,in=-60,looseness=7] (a);
      \draw [ultra thick,->>, blue](b) to [out=-60,in=0,looseness=5] (b);
     \draw [ultra thick,->>, blue](c) to [out=180,in=240,looseness=5] (c);

    \node (a1) at (1.8,0) {$\mathbf{(4)}$}; 
     \node (a2) at (2.3,.866) {$\mathbf{(1)}$};
     \node (a3) at (2.8,1.732) {$\mathbf{(-2)}$};
     \path[ultra thick,<<-](a1) edge node {} (a2);
     
      \draw [ultra thick,->>,blue](a1) to [out=-120, in=-60,looseness=7] (a1);
      \path[ultra thick,<<-](a2) edge node {} (a3);
       \draw [ultra thick,->>, blue](a3) to [out=60, in=120,looseness=7] (a3);
        \draw [ultra thick,->>, blue](a2) to [out=-60, in=0,looseness=5] (a2);
     
      \node (b1) at (1.4,.866) {};
      \node (b2) at (2,.866) {};
       \node (b3) at (.9,0) {};
      \node (b4) at (1.5,0) {};
       \node (b5) at (1.9,1.732) {};
      \node (b6) at (2.5,1.732) {};
      
      \path[dotted, thick](b1) edge  node [below]{} (b2);
      \path[dotted, thick](b3) edge  node [below]{} (b4);
      \path[dotted, thick](b5) edge  node [below]{} (b6);

\end{tikzpicture}
\end{center}
\caption{Quiver diagram for the representation $\underline{\mathbf{6}}$ with the same notation as before.}
\label{fig:6}
\end{figure}
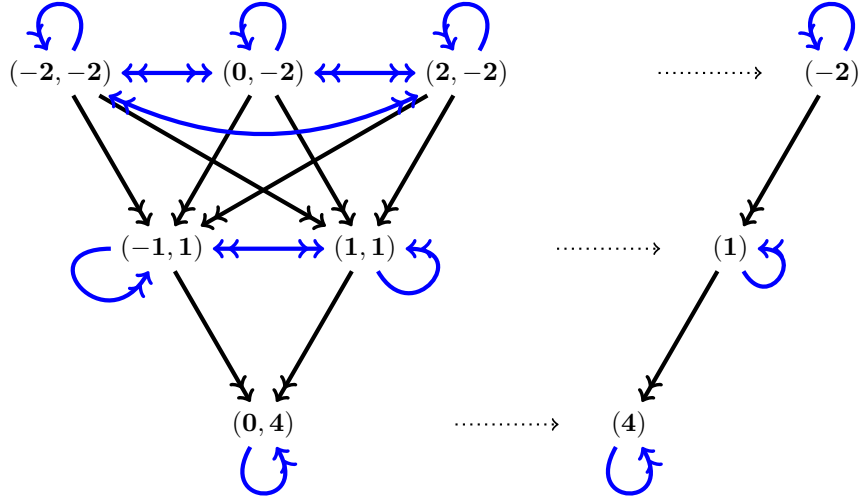

% \item {{\bf Representation $\underline{\mathbf{6}}$:} 
\paragraph{Representation $\underline{\mathbf{6}}$}
The six-dimensional representation $\underline{\mathbf{6}}$ of $\mathrm{SU}(3)$ causes 
the more complicated quiver 
diagram depicted in \mbox{Figure \ref{fig:6}},
which could be also obtained by the direct evaluation of  the commutation relation with the
$\uo_{1,1}$-generator $\hat{I}_8 =  \mathrm{diag}\lb 4,1,1,-2,-2,-2\rb$. The resulting fields are of the form
\small
\bea
\phi^{(1)} \ \text{and}\ \phi^{(2)\+} &=& \begin{pmatrix}
                       0 & \ast &\ast  & 0 & 0 &0\\
                       0 & 0    & 0    & \ast & \ast & \ast\\
                       0 & 0    & 0    & \ast & \ast & \ast\\
                       0 & 0 & 0 & 0 & 0 & 0\\
                       0 & 0 & 0 & 0 & 0 & 0\\
                       0 & 0 & 0 & 0 & 0 & 0
                      \end{pmatrix}, \ \ 
\phi^{(3)} \ \text{and} \ X_7 \= \begin{pmatrix}
                             \ast& 0& 0& 0 &0 & 0\\
                             0& \ast & \ast & 0 &0 & 0\\
                             0& \ast & \ast & 0 &0 & 0\\
                             0& 0 & 0& \ast & \ast & \ast\\
                             0& 0 & 0& \ast & \ast & \ast\\ 
                             0& 0 & 0& \ast & \ast & \ast
                            \end{pmatrix}.
\eea
\normalsize 
We skip the explicit index structure of the invariant gauge connection which can be read from
the quiver diagram, Figure~\ref{fig:6}, and provide only the result for the modified holomorphic chain 
\bea
\label{eq:conn_6}
{\ca}_{\underline{\mathbf{6}}} \= \begin{pmatrix}
                             \frac{4}{\sqrt{3}} \im e^8 \otimes \id +\Psi_{4;4} &  \tilde{\Phi}_{1;4}&0\\
                             -\tilde{\Phi}_{1;4}^{\+} & \frac{1}{\sqrt{3}} \im e^8 \otimes \tilde{\id} +\tilde{\Psi}_{1;1} &  \tilde{\Phi}_{-2;1} \\
                             0      & -\tilde{\Phi}_{-2;1}^{\+}    & -\frac{2}{\sqrt{3}} \im e^8 \otimes \tilde{\id} +\tilde{\Psi}_{-2;-2}
                            \end{pmatrix}.
\eea
It is interesting to compare this block matrix  of size $3\times 3$ with that of the adjoint representation in the last example, which -- on the level 
of the modified holomorphic chain -- only differs in
the occurring quantum numbers and, thus, the connection $\Gamma$.

% }

\begin{figure}[t!]
\begin{center}
\small
    \begin{tikzpicture}[->,scale=2.7]
    \node (a) at (.5,-.866) {$\mathbf{(1,3)}$};
    \node (b) at (-.5,-.866) {$\mathbf{(-1,3)}$};
    \node (c) at (1,0) {$\mathbf{(2,0)}$};
    \node (d) at (0,0) {$\mathbf{(0,0)^2}$};
    \node (e) at (-1,0) {$\mathbf{(-2.0)}$};
    \node (f) at (.5,.866) {$\mathbf{(1,-3)}$};
    \node (g) at (-.5,.866) {$\mathbf{(-1,-3)}$};

     \path[ultra thick,<<-](a) edge node {} (c);
     \path[thick,<<-](a) edge [double] node {} (d);
     \path[ultra thick,<<-](a) edge  node {} (e);
     
     \path[ultra thick,<<-](b) edge node {} (c);
     \path[thick,<<-](b) edge [double] node {} (d);
     \path[ultra thick,<<-](b) edge node {} (e);
     
     \path[ultra thick,<<-](c) edge node {} (f);
     \path[ultra thick,<<-](c) edge node {} (g);
     \path[thick,<<-](d) edge [double] node {} (f);
     \path[thick,<<-](d) edge [double] node {} (g);
     \path[ultra thick,<<-](e) edge node {} (f);
     \path[ultra thick,<<-](e) edge node {} (g);

     \path[<<->>,ultra thick,  blue](b) edge node {} (a);
     \path[<<->>,thick,  blue](e) edge [double] node {} (d);
     \path[<<->>,thick,  blue](d) edge [double] node {} (c);
     \path[<<->>,ultra thick,  blue](e) edge [bend left=24] node {} (c);
     \path[<<->>,ultra thick,  blue](g) edge node {} (f);
 
     \draw [ultra thick,->>, blue](f) to [out=60,in=120,looseness=7] (f);
     \draw [ultra thick,->>, blue](g) to [out=60,in=120,looseness=7] (g);
     \draw [ultra thick,->>, blue](a) to [out=-120,in=-60,looseness=7] (a);
     \draw [ultra thick,->>, blue](b) to [out=-120,in=-60,looseness=7] (b);
     \draw [thick,->>, blue](d) to [out=-110,in=-70,looseness=12,double] (d);
     \draw [ultra thick,->>, blue](c) to [out=-30,in=30,looseness=4] (c);
     \draw [ultra thick,->>, blue](e) to [out=150,in=210,looseness=4] (e);

     \node (a1) at (2,-0.866) {$\mathbf{(3)}$}; 
     \node (a2) at (2.5,0) {$\mathbf{(0)}$};
     \node (a3) at (3,.866) {$\mathbf{(-3)}$};
     \path[ultra  thick,<<-](a1) edge node {} (a2);
     
     \draw [ultra thick,->>, blue](a1) to [out=-120, in=-60,looseness=7] (a1);
     \path[ultra thick,<<-](a2) edge node {} (a3);
     \draw [ultra thick,->>, blue](a3) to [out=60, in=120,looseness=7] (a3);
     \draw [ultra thick,->>, blue](a2) to [out=-60, in=0,looseness=5] (a2);
     
     \node (b1) at (.8,-.866) {};
     \node (b2) at (1.7,-.866) {};
     \node (b3) at (1.7,0) {};
     \node (b4) at (2.2,0) {};
     \node (b5) at (1.8,.866) {};
     \node (b6) at (2.7,.866) {};
      
      \path[dotted, thick](b1) edge  node [below]{} (b2);
      \path[dotted, thick](b3) edge  node [below]{} (b4);
      \path[dotted, thick](b5) edge  node [below]{} (b6);
\end{tikzpicture}
\end{center}
\caption{Quiver diagram for the adjoint representation $\underline{\mathbf{8}}$ of $\mathrm{SU}(3)$. Note that due to the degeneracy of $(0,0)$ 
each arrow involving 
the origin must be counted twice (depicted as arrows consisting of two lines), i.e. there are, for instance, \emph{four}
 arrows between $(0,0)$ and $(1,-3)$ etc.}
\label{fig:adj}
\end{figure}
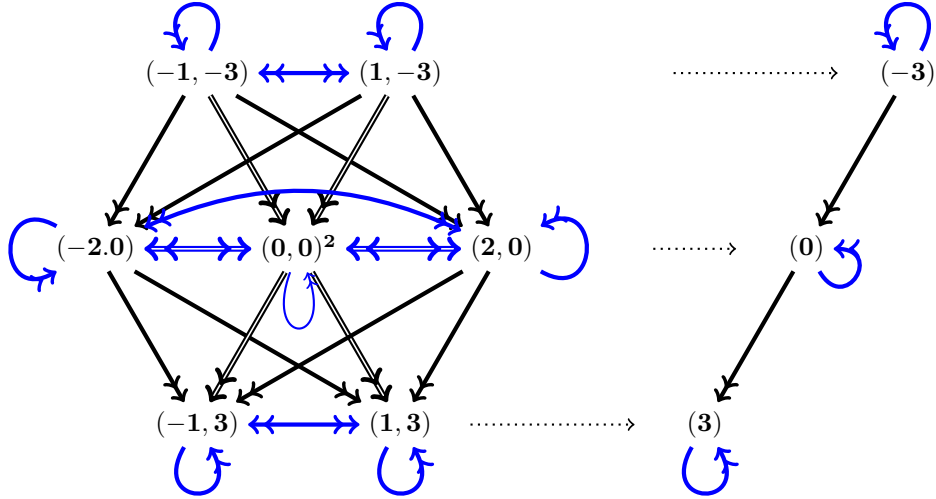

% \item{{\bf Adjoint representation $\underline{\mathbf{8}}$:} 
\paragraph{Adjoint representation $\underline{\mathbf{8}}$}
The $\mathrm{U}(1)_{1,1}$-generator 
in the adjoint representation is given by $\hat{I}_8= \mathrm{diag}\lb 3,3,0,0,0,0,-3,-3\rb$ and the weight diagram is a hexagon with
 two degenerated points at the origin\footnote{The representation of the other Cartan generator reads
$\mathrm{ad}(-\im I_7)=\mathrm{diag}\lb -1,1, -2,0,0,2,-1,1\rb$, which causes the degeneracy at $(0,0)$.}. The Higgs fields must thus have the shape
\footnotesize
\bea
\phi^{(1)} \ \text{and} \ \phi^{(2)\+} = 
\begin{pmatrix}
0	&0	&\ast	&\ast	&\ast	&\ast	&0	&0\\
0	&0	&\ast	&\ast	&\ast	&\ast	&0	&0\\
0	&0	&0	&0	&0	&0	&\ast	&\ast\\
0	&0	&0	&0	&0	&0	&\ast	&\ast\\ 
0	&0	&0	&0	&0	&0	&\ast	&\ast\\ 
0	&0	&0	&0	&0	&0	&\ast	&\ast\\
0	&0	&0	&0	&0	&0	&0	&0\\   
0	&0	&0	&0	&0	&0	&0	&0                                                   
\end{pmatrix}, \ 
 \phi^{(3)} \ \text{and} \ X_7 =
 \begin{pmatrix}
\ast	&\ast	&0	&0	&0	&0	&0	&0\\
\ast	&\ast	&0	&0	&0	&0	&0	&0\\
0	&0	&\ast	&\ast	&\ast	&\ast	&0	&0\\
0	&0	&\ast	&\ast	&\ast	&\ast	&0	&0\\
0	&0	&\ast	&\ast	&\ast	&\ast	&0	&0\\
0	&0	&\ast	&\ast	&\ast	&\ast	&0	&0\\
0	&0	&0	&0	&0	&0	&\ast	&\ast\\  
0	&0	&0	&0	&0	&0	&\ast	&\ast                                                   
\end{pmatrix},
\eea
\normalsize
and the quiver diagram Figure~\ref{fig:adj} contains a large number of arrows. 
The identification leading to the modified holomorphic chain yields as connection
\bea
{\ca}_{\underline{\mathbf{6}}} \= \begin{pmatrix}
                             \sqrt{3} \im e^8 \otimes \id +\Psi_{3;3} & \tilde{\Phi}_{0;3}&0\\
                             -\tilde{\Phi}_{0;3}^{\+} &  \tilde{\Psi}_{0;0} &  \tilde{\Phi}_{-3;0} \\
                             0      & -\tilde{\Phi}_{-3;0}^{\+}    & -\sqrt{3} \im e^8 \otimes \tilde{\id} +\tilde{\Psi}_{-3;-3}
                            \end{pmatrix}.
\eea
As mentioned before, this modified holomorphic chain of length 3 is different from that of the six-dimensional representation, (\ref{eq:conn_6}),
only due to the quantum numbers that appear.
% }
% \end{itemize}

The huge number of arrows in the last two examples have shown that it is advantageous to use only the relevant quantum number 
$\n_8$ rather than the entire weight diagram of $G$, but for comparisons with $Q_3$ the latter description is also useful.
The occurrence of degeneracies in the entire weight diagram of $\mathrm{SU}(3)$ due to the weaker equivariance condition is similar \cite{LPS06, T11} to 
the  case of the five-dimenional Sasaki-Einstein manifold $T^{1,1}\coloneqq \lb \mathrm{SU}(2)\times \mathrm{SU}(2)\rb/\mathrm{U}(1)$ 
in comparison with its underlying manifold $\C P^1 \times \C P^1$. 

%=========================================================
%=========================================================
%=========================================================
%=========================================================
%=========================================================
%=========================================================
\subsection{Dimensional reduction of the Yang-Mills action}

In the previous section we have completely characterized the form of a $G$-invariant gauge connection by applying 
the rules (\ref{eq:eq_rules_aw}) in the weight diagram and in terms of the results (\ref{eq:eff_conn1}) and (\ref{eq:eff_conn2}).
Given such a gauge connection $\ca$ on $M^d \times X_{1,1}$ with field strength $\cf$, we now determine its  standard Yang-Mills action 
\bea
\label{eq:YM_action}
S_{\mathrm{YM}} \= -\frac{1}{4} \int_{M^d \times X_{1,1}} \mathrm{tr} \ \cf \wedge \ast \cf,
\eea
yielding the usual Yang-Mills Lagrangian\footnote{We use the set of indices  $\left\{\hat{\m} \right\}= \left\{\m, \a, \bar{\a},7\right\}$ with 
 $\m$ referring to  $M^d$.} 
\bea
\mathcal{L}_{\mathrm{YM}}\= - \frac{1}{4} \sqrt{\hat{g}}\ \mathrm{tr}\ \cf_{\hat{\m}\hat{\n}} \cf^{\hat{\m}\hat{\n}}, 
\eea 
where we denote $\hat{g} \coloneqq \mathrm{det}\ g_{X_{1,1}} \mathrm{det}\ g_{M^d}$. Using the Sasaki-Einstein metric (\ref{eq:metric_aw}),
\bea
\lb g_{X_{1,1}} \rb_{\a \bar{\b}} \= \frac{1}{2} \delta_{\a\b}\ \ \ \ \ \text{and} \ \ \ \ \ \lb g_{X_{1,1}}\rb_{77}\=1, 
\eea
and the field strength components from (\ref{eq:fieldstrength}), one obtains as Lagrangian
\small
\bea
\label{eq:lagr_aw}
\nonumber  \mathcal{L}_{\mathrm{YM}} &=& \sqrt{\hat{g}}\ \mathrm{tr}_k \left\{ \frac{1}{4} F_{\m\n} \lb F^{\m \n}\rb^{\+} 
   +2 \sum_{\a=1}^3 \left| D_{\m}\phi^{(\a)} \right|^2 +\frac{1}{2} \left| D_{\m} X_7\right|^2  
     +2 \left|\com{\phi^{(1)}}{\phi^{(1)\+}} - \im X_7 +\sqrt{3} \im I_8 \right|^2 \right.\\
\nonumber &&+ \left.  2 \left|\com{\phi^{(2)}}{\phi^{(2)\+}} - \im X_7 -\sqrt{3}\im I_8 \right|^2
     +2 \left|\com{\phi^{(3)}}{\phi^{(3)\+}} - \im X_7\right|^2 + 4 \left|\com{\phi^{(1)}}{\phi^{(2)}}-2 \phi^{(3)}\right|^2 \right.\\
\nonumber  &&+ \left.   4 \left|\com{\phi^{(1)}}{\phi^{(3)}} \right|^2 + 4 \left|\com{\phi^{(2)}}{\phi^{(3)}} \right|^2
        +4 \left|\com{\phi^{(1)}}{\phi^{(2)\+}}\right|^2  + 4 \left| \com{\phi^{(1)}}{\phi^{(3)\+}} + \phi^{(2)\+}\right|^2 \right.\\
\nonumber && +\left. 4 \left| \com{\phi^{(2)}}{\phi^{(3)\+}}- \phi^{(1)\+}\right|^2 +2 \left| \com{\phi^{(1)}}{X_7}-\im \phi^{(1)}\right|^2 
+ 2 \left| \com{\phi^{(2)}}{X_7}-\im \phi^{(2)}\right|^2 \right.\\ 
 &&  + 2 \left| \com{\phi^{(3)}}{X_7}-2 \im \phi^{(3)} \right|^2 \Bigg\}.
\eea
\normalsize
Here, we have defined the covariant derivatives $D_{\m} \phi^{(\a)} \coloneqq \lb \diff \phi^{(\a)} +\com{A}{\phi^{(\a)}}\rb_{\m}$ for $\a=1, 2, 3$
and $D_{\m} X_7 \coloneqq \lb \diff X_7 +\com{A}{X_7}\rb_{\m}$, the field strength
$F_{\m\n}\coloneqq \lb \diff A +A\wedge A\rb_{\m\n}$ and we write $|X|^2 \coloneqq X X^{\+}$.
Since the fields $\phi^{(\a)}$ and $X_7$ are assumed to be independent from internal coordinates of $X_{1,1}$ (due to equivariance), the additional 
dimensions can be integrated out easily, which yields only a prefactor $\mathrm{vol}\lb X_{1,1}\rb$ for the dimensional reduction of the Lagrangian. 
In this way, one obtains  from a pure Yang-Mills theory on $M^d\times X_{1,1}$ a Yang-Mills-Higgs action on  $M^d$, where 
the endomorphisms $\phi^{(a)}$ and $X_7$ constitute a non-trivial potential provided by the internal geometry of $X_{1,1}$.
%=========================================================
%=========================================================
%=========================================================
%=========================================================
%=========================================================
%=========================================================
\subsection{Reduction to quiver gauge theory on $Q_3$}

\begin{figure}[t!]
\begin{center}
    \begin{tikzpicture}[->,scale=2.5]
    \node (a0) at (-2.6,.5) {$\mathbf{a)}$};
    \node (a) at (-1.8,0) {$\mathbf{(0,2)}$};
    \node (b) at (-1.3,.866) {$\mathbf{(1,-1)}$};
    \node (c) at (-2.3,.866) {$\mathbf{(-1,-1)}$};

    \path[ultra thick](b) edge  node {} (a);
    \path[ultra thick, red](a) edge  node {} (c);
    \path[->,ultra thick,  blue](b) edge  node {} (c);

    \node (a2) at (.5,-2.666) {$\mathbf{(1,3)}$};
    \node (b2) at (-.5,-2.666) {$\mathbf{(-1,3)}$};
    \node (c2) at (1,-1.8) {$\mathbf{(2,0)}$};
    \node (d2) at (0,-1.8) {$\mathbf{(0,0)^2}$};

    \node (e2) at (-1,-1.8) {$\mathbf{(-2,0)}$};
    \node (f2) at (.5,-.934) {$\mathbf{(1,-3)}$};
    \node (g2) at (-.5,-.934) {$\mathbf{(-1,-3)}$};
    
    \node (a02) at (-1,-1.3) {$\mathbf{c)}$};
    
    \path[ultra thick](c2) edge  node {} (a2);
    \path[thick](d2) edge  [double] node {} (b2);
    \path[thick](f2) edge  [double] node {} (d2);
    \path[ultra thick](g2) edge  node {} (e2);
          
    \path[thick, red](a2) edge  [double] node {} (d2);
    \path[ultra thick, red](b2) edge  node {} (e2);
    \path[ultra thick, red](c2) edge  node {} (f2);
    \path[thick, red](d2) edge  [double] node {} (g2);      
          
  \path[->,ultra thick,  blue](a2) edge  node {} (b2);
  \path[->,thick,  blue](c2) edge  [double] node {} (d2);
  \path[->,thick,  blue](d2) edge  [double] node {} (e2);
  \path[->,ultra thick, blue](f2) edge  node {} (g2);

    \node (a03) at (1.3,-.5) {$\mathbf{b)}$};    
    \node (a3) at (1.8,-.866) {$\mathbf{(0,4)}$};
    \node (b3) at (2.3,0) {$\mathbf{(1,1)}$};
    \node (c3) at (1.3,0) {$\mathbf{(-1,1)}$};
    \node (d3) at (2.8,.866) {$\mathbf{(2,-2)}$};
    \node (e3) at (1.8,.866) {$\mathbf{(0,-2)}$};
    \node (f3) at (.8,.866) {$\mathbf{(-2,-2)}$};  
    
    \path[ultra thick](b3) edge  node {} (a3);
    \path[ultra thick](d3) edge  node {} (b3);
    \path[ultra thick](e3) edge  node {} (c3);
    
    \path[ultra thick, red](a3) edge  node {} (c3);
    \path[ultra thick, red](c3) edge  node {} (f3);
    \path[ultra thick, red](b3) edge  node {} (e3);
       
    \path[->,ultra thick,  blue](b3) edge  node {} (c3);
    \path[->,ultra thick,  blue](d3) edge  node {} (e3);
    \path[->,ultra thick,  blue](e3) edge  node {} (f3); 
   
\end{tikzpicture}
\end{center}
\caption{Quiver diagrams of $Q_3$ for a) fundamental representation $\underline{\mathbf{3}}$, b) representation $\underline{\mathbf{6}}$, and c) 
adjoint representation $\underline{\mathbf{8}}$ (with the degenerated origin) of $\mathrm{SU}(3)$. The arrows denote the Higgs 
fields $\phi^{(1)}$ (black), $\phi^{(2)}$ (red), and
$\phi^{(3)}$ (blue), according to the condition (\ref{eq:eq_rules_q3}).}
\label{fig:q3}
\end{figure}
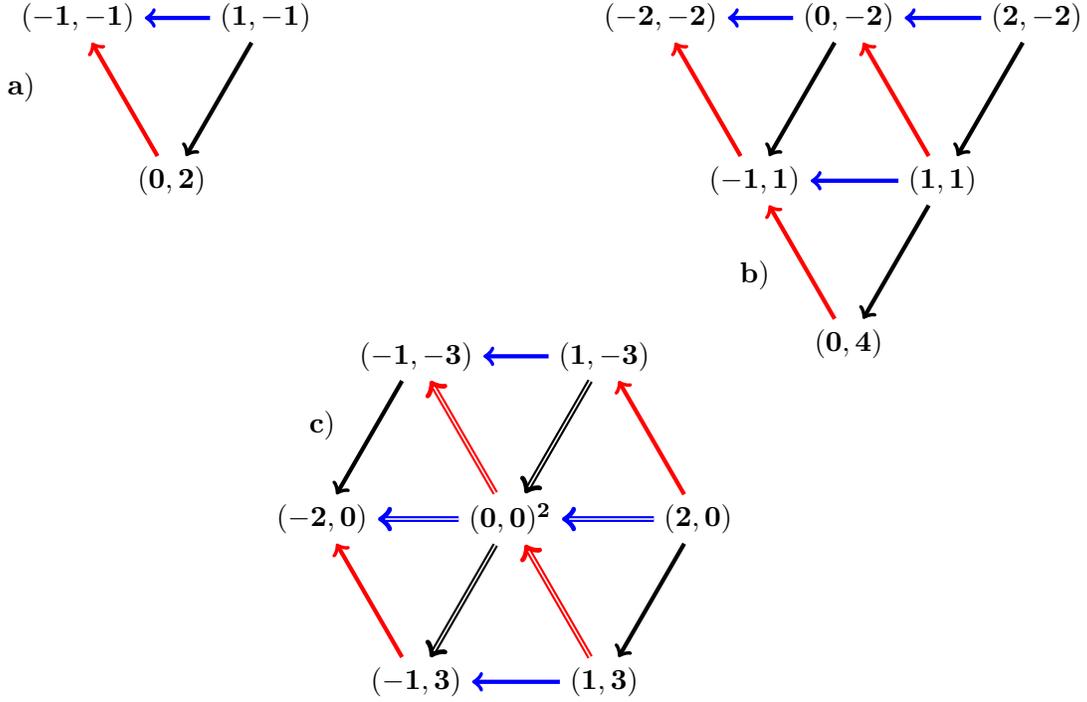

The equivariance condition and the examples of the quiver diagrams in the previous section have shown that the quiver gauge theory on $X_{1,1}$ 
depends on only one of the two quantum numbers of $\mathrm{SU}(3)$. This yields effectively a modified holomorphic chain as quiver diagram or, 
considered in the 
original weight diagram of $\mathrm{SU}(3)$, a diagram with multiple arrows and degeneracies.
As mentioned in the discussion of the Sasaki-Einstein structure on $X_{1,1}$ in Section \ref{sec:sasaki_einstein}, the space is a $\mathrm{U}(1)$-bundle over 
the (K\"ahler) space $Q_3$, so that it is natural to consider 
the reduction from the gauge theory on  $X_{1,1}$ to that on $Q_3$ by removing the 
contact direction as a degree of freedom. Since we then divide by a Cartan subalgebra, the quiver diagram is simply the weight diagram 
of $\mathrm{SU}(3)$ without the degeneracies which have been caused 
by the weaker conditions on $X_{1,1}$. This reduction can be performed by setting the terms containing
$e^7 \wedge \Theta^{\a}$ or $e^7 \wedge \Theta^{\bar{\a}}$ in the field strength (\ref{eq:fieldstrength}) to zero. This provides the additional
equivariance conditions
\bea
\label{eq_:eq_q3}
\com{X_7}{\phi^{(1)}} &=& -\im \phi^{(1)}, \ \ \ \ \com{X_7}{\phi^{(2)}}\= -\im \phi^{(2)}, \ \ \text{and} \ \ \ \com{X_7}{\phi^{(3)}} \= -2\im \phi^{(3)}. 
\eea
For the reduction to $Q_3$, the field $X_7$ must be proportional to $I_7$ and setting $X_7= I_7$ fixes the action of the Higgs fields to be
\bea
\label{eq:eq_rules_q3}
\nonumber &&\phi^{(1)}: \lb \n_7,\n_8\rb \longmapsto (\n_7-1,\n_8+3)\\
          &&\phi^{(2)}: \lb \n_7,\n_8 \rb \longmapsto (\n_7-1,\n_8-3)\\
\nonumber &&\phi^{(3)}: \lb \n_7,\n_8\rb \longmapsto (\n_7-2,\n_8).
\eea
This, indeed, requires the quiver diagrams in Figure \ref{fig:q3} to coincide with the weight diagrams of the chosen representations and yields the 
results\footnote{Note that the orientation of the Higgs fields depends on the chosen convention of the holomorphic structure; we 
denote as Higgs fields $\phi^{\a}$ the endomorphisms accompanying the anti-holomorphic forms $\Theta^{\bar{\a}}$.} 
from \cite{SU3,double_quiver}. The endomorphism part of the gauge connection, e.g. 
for the fundamental representation, reads
\bea
\ca_{\underline{\mathbf{3}}} &=& \begin{pmatrix}
         \id \otimes  \frac{2}{\sqrt{3}} \im e^8  & - \Phi^{(2)\+}_{0,2;-1,-1}   &  \Phi^{(1)}_{1,-1;0,2}\\
           \Phi^{(2)}_{0,2;-1,-1} & \id\otimes\lb -\frac{1}{\sqrt{3}} \im e^8 - \im e^7\rb   &  \Phi^{(3)}_{1,-1;-1,-1} \\
          -\Phi^{(1)\+}_{1,-1;0,2} & -\Phi^{(3)\+}_{1,-1;-1,-1} &\id \otimes \lb -\frac{1}{\sqrt{3}} \im e^8 +\im e^7\rb  
        \end{pmatrix}
\eea
with $\Phi^{(\a)}\coloneqq \phi^{(\a)} \otimes \Theta^{\bar{\a}}$. Since the quiver diagram is the weight diagram of $\mathrm{SU}(3)$, the Higgs fields 
have the block shape of the generators (\ref{eq:generators}) and the central idea of quiver gauge theory becomes evident: 
One modifies the  bundle (\ref{eq:def_1forms}) by inserting compatible endomorphisms $\phi^{(\a)}$ as entries in the block matrices 
describing the gauge connection.

The Lagrangian of the gauge theory
on $M^d \times Q_3$ is then  given by that on $M^d \times X_{1,1}$ without the terms containing commutators \mbox{with $X_7$},
\small
\bea
\label{eq:lagr_q3}
\nonumber  \mathcal{L}_{Q_3} &=& \sqrt{\hat{g}}\ \mathrm{tr}_k \left\{ \frac{1}{4} F_{\m\n} \lb F^{\m \n}\rb^{\+} 
   +2 \sum_{\a=1}^3 \left| D_{\m}\phi^{(\a)} \right|^2  
     +2 \left|\com{\phi^{(1)}}{\phi^{(1)\+}} -\im I_7 +\sqrt{3} \im I_8 \right|^2 \right. \\
\nonumber && + \left.  2 \left|\com{\phi^{(2)}}{\phi^{(2)\+}} - \im I_7 -\sqrt{3}\im I_8 \right|^2
     +2 \left|\com{\phi^{(3)}}{\phi^{(3)\+}} -\im I_7\right|^2 + 4 \left|\com{\phi^{(1)}}{\phi^{(2)}}-2 \phi^{(3)}\right|^2 \right.\\
\nonumber  &&+ \left.   4 \left|\com{\phi^{(1)}}{\phi^{(3)}} \right|^2 + 4 \left|\com{\phi^{(2)}}{\phi^{(3)}} \right|^2
        +4 \left|\com{\phi^{(1)}}{\phi^{(2)\+}}\right|^2  + 4 \left| \com{\phi^{(1)}}{\phi^{(3)\+}} + \phi^{(2)\+}\right|^2 \right.\\
 && + 4 \left| \com{\phi^{(2)}}{\phi^{(3)\+}}- \phi^{(1)\+}\right|^2 \Bigg\} ,
\eea
\normalsize
because the vanishing of them is subject to the  further equivariance conditions (\ref{eq_:eq_q3}).

%=========================================================
%=========================================================
%=========================================================
%=========================================================
%=========================================================
%=========================================================
\section{Instantons on the metric cone $C\lb X_{1,1}\rb $}
\label{sec:metric_cone}
The implementation of the equivariance condition (\ref{eq:equivariance_aw}) has determined the general 
form of the gauge connection, expressed in the associated quiver diagram, and the action functional, but has not restricted the entries of 
the endomorphisms. Further conditions and relations among the endomorphisms can be imposed by studying vacua of the gauge theory, i.e. by minimizing the action
functional (\ref{eq:YM_action}).  To this end, we will evaluate the \emph{Hermitian Yang-Mills equations} -- a certain form of  
generalized self-duality equations -- on the metric cone $C(X_{1,1})$, 
as it has been done in similar setups, e.g. \cite{S5,T11}, and describe their moduli space, following \cite{Donaldson84, Kronheimer90,MS15}. 
 %=========================================================
%=========================================================
%=========================================================
%=========================================================
%=========================================================
%=========================================================
\subsection{Generalized self-duality equation}
A very useful tool for obtaining minima of a Yang-Mills functional  in gauge theory is to evaluate a first-order equation implying 
the second-order Yang-Mills equations \cite{Corrigan83, Ward84, Hull}. Given a connection $\ca$ on an $n$-dimensional manifold whose 
curvature $\cf$ satisfies the generalized self-duality equation
\bea
\label{eq:inst}
\ast \cf \= -\ast Q \wedge \cf
\eea
for a \mbox{4-form} $Q$, one obtains by taking the differential \cite{noelle12}
\bea
\label{eq:YMT}
\nabla^{\ca} \wedge \ast \cf + \lb \diff \ast Q\rb \wedge \cf &=& 0  \  \ \text{with} \ \
\nabla^{\ca} \wedge \ast \cf \coloneqq \diff \ast \cf + \ca \wedge \ast \cf + \lb -1\rb^{n-1} \ast \cf \wedge \ca,
\eea
\normalsize
which is the usual Yang-Mills equation with torsion term $\lb \diff \ast Q\rb \wedge \cf$. Explicit formulae for the choice of the form $Q$, 
in dependence of the geometry of the manifold, such that the torsion term vanishes even if the form $Q$ is not co-closed have been given in \cite{noelle12}.
Their construction is based on the existence of (real) Killing spinors, and thus also applies 
to Sasaki-Einstein structures. A connection $\ca$ whose curvature satisfies (\ref{eq:inst}) for the form $Q$ given by \cite{noelle12}
is called a \emph{(generalized) instanton}. For a Sasaki-Einstein manifold the form $Q$ reads \cite{noelle12}
\bea
Q\= \frac{1}{2} \omega \wedge \omega,
\eea
such that we have 
\bea
\label{eq:Q_SE}
Q \= -\frac{1}{4} \lb \Theta^{1\bar{1}2\bar{2}}+\Theta^{1\bar{1}3\bar{3}}+\Theta^{2\bar{2}3\bar{3}}\rb \= e^{1234}+e^{1256}+e^{3456}.
\eea
The corresponding instanton equation (\ref{eq:inst}) on $X_{1,1}$ is solved by the connection $\Gamma= I_8 \otimes e^8$, which we used for expressing 
the $G$-invariant connection in 
(\ref{eq:inv_conn}); see  Appendix \ref{sec:connections}. The form $Q_Z$ occurring in the instanton equation on the cylinder, which 
is conformally equivalent to the metric cone, 
over a Sasaki-Einstein manifold 
reads \cite{noelle12}\footnote{They provide the form for a whole familiy of compatible metrics and we consider one special value here.}
\bea
Q_Z \= \diff \tau \wedge P + Q \ \ \ \ \ \text{with} \ \ \ \ \ P\= \eta \wedge \omega
\eea
and one thus obtains
\bea
Q_Z \= \frac{1}{2} \Omega \wedge \Omega,
\eea
where $\Omega$ is the K\"ahler form of the Calabi-Yau cone and the cylinder, respectively. 
Since the Calabi-Yau manifold is of complex dimension 4 and as we have chosen the standard form  of the K\"ahler form, the \mbox{4-form}
 $Q_Z$ is self-dual, such that $\diff \ast Q_Z = \diff Q_Z =0$, and  the Yang-Mills equation without torsion  follows from 
the instanton equation (\ref{eq:inst}). 
We evaluate the instanton equation  (\ref{eq:inst}) with the form $Q_Z$ by imposing the (equivalent) \emph{Hermitian Yang-Mills equations} 
(HYM) \cite{Popov09, Donaldson85, UY86}
\bea
\label{eq:HYM}
\cf^{(2,0)}\= 0 \= \cf^{(0,2)} \ \ \ \ \ \ \ \text{and} \ \ \ \ \ \ \ \Omega \haken \cf \coloneqq \ast \lb \Omega \wedge \ast \cf  \rb \=0,
\eea 
where $\cf^{(2,0)}$ refers to the $(2,0)$-part with respect to the complex \mbox{structure $J$}. The first 
equation is a holomorphicity condition and the second one can (sometimes) be considered as a stability 
condition on  vector bundles; they are also known as \emph{Donaldson-Uhlenbeck-Yau equations}.

%=========================================================
%=========================================================
%=========================================================
%=========================================================
%=========================================================
%=========================================================

\subsection{Hermitian Yang-Mills instantons on $C\lb X_{1,1}\rb$}
We consider the same ansatz (\ref{eq:conn}) \cite{AW11}, now including also the additional form $e^{\tau}\coloneqq  \diff \tau \coloneqq \frac{\diff r}{r}$
on the cylinder, 
\bea
\label{eq:ansatz_cone}
\nonumber \ca &=& I_8 e^8  +Y_{\a} \Theta^{\a} + Y_{\bar{\a}} \Theta^{\bar{\a}} + X_7 e^7 +X_{\tau} e^{\tau}\\
             &=& I_8 e^8 +Y_{\a} \Theta^{\a} + Y_{\bar{\a}} \Theta^{\bar{\a}} + Y_4 \Theta^4 + Y_{\bar{4}} \Theta^{\bar{4}},
\eea
where we set\footnote{The field $X_{\tau}$ associated to
the radial direction could be gauged to zero \cite{AW11}.}
\bea
Y_4 \coloneqq \frac{1}{2} \lb X_{\tau}+ \im X_7\rb.
\eea 
Due to the equivariance, the endomorphisms are ``spherically symmetric'', $X_a = X_a \lb r\rb$.  
After the implementation of  the same equivariance conditions as before,
\bea
\label{eq:equivariance}
\com{I_8}{Y_{\bar{1}}}&=& -\sqrt{3} \im Y_{\bar{1}}, \ \ \ \com{I_8}{Y_{\bar{2}}}\= \sqrt{3} \im Y_{\bar{2}}, \ \ \ \text{and} \ \ \ 
\com{I_8}{Y_{\bar{3}}} \= 0 \= \com{I_8}{Y_{\bar{4}}},
\eea
the non-vanishing components of the field strength read
\bea
\nonumber \cf_{\a\b} &=& \com{Y_{\a}}{Y_{\b}} -C_{\a \b}^{\g} Y_{\g}, \ \ \ \cf_{\bar{\a}\bar{\b}}\= \com{Y_{\bar{\a}}}{Y_{\bar{\b}}}- 
C_{\bar{\a}\bar{\b}}^{\bar{\g}} Y_{\bar{\g}},\\
\nonumber \cf_{\a \bar{\b}} &=& \com{Y_{\a}}{Y_{\bar{\b}}} -C_{\a\bar{\b}}^{\g}Y_{\g} - C_{\a \bar{\b}}^{\bar{\g}} Y_{\bar{\g}}
+ C_{\a\bar{\b}}^7 Y_4 - C_{\a \bar{\b}}^7 Y_{\bar{4}} +\im C_{\a \bar{\b}}^8 I_8,\\
\cf_{\a 4} &=& \com{Y_{\a}}{Y_4}- \frac{1}{2}r \dot{Y}_{\a} - \frac{1}{2} C_{7\a}^{\b}Y_{\b}, \ \ \ 
\cf_{\a \bar{4}} \= \com{Y_{\a}}{Y_{\bar{4}}}- \frac{1}{2}r \dot{Y}_{\a} + \frac{1}{2} C_{7\a}^{\b}Y_{\b},\\
\nonumber \cf_{\bar{\a}4} &=& \com{Y_{\bar{\a}}}{Y_4} -\frac{1}{2}r \dot{Y}_{\bar{\a}} - \frac{1}{2} C_{7\bar{\a}}^{\bar{\b}}Y_{\bar{\b}}, \ \ \
\cf_{\bar{\a}\bar{4}} \= \com{Y_{\bar{\a}}}{Y_{\bar{4}}} -\frac{1}{2}r \dot{Y}_{\bar{\a}} + \frac{1}{2} C_{7\bar{\a}}^{\bar{\b}}Y_{\bar{\b}},\\
\nonumber \cf_{4\bar{4}} &=& \com{Y_4}{Y_{\bar{4}}} -\frac{1}{2}r \lb \dot{Y}_4-\dot{Y}_{\bar{4}}\rb.
\eea
Evaluating the condition $\cf_{\bar{\a}\bar{\b}}=0$ leads to 
\bea
\label{eq:constraints}
\com{Y_{\bar{1}}}{Y_{\bar{2}}}&=& 2 Y_{\bar{3}}, \ \ \ \com{Y_{\bar{1}}}{Y_{\bar{3}}} \= 0 \= \com{Y_{\bar{2}}}{Y_{\bar{3}}}
\eea
(together with their complex conjugates from $\cf_{\a\b}=0$). Thus, this part of the holomorphicity condition imposes algebraic relations on the quiver.
In contrast, from $\cf_{\bar{\a}\bar{4}}=0$ we obtain the following flow equations
\bea
\label{eq:flow1}
r \dot{Y}_{\bar{1}} &=& -Y_{\bar{1}} +2 \com{Y_{\bar{1}}}{Y_{\bar{4}}}, \ \ \ \ 
r \dot{Y}_{\bar{2}} \= -Y_{\bar{2}} +2 \com{Y_{\bar{2}}}{Y_{\bar{4}}}, \ \ \ \
 r \dot{Y}_{\bar{3}} \= -2 Y_{\bar{3}} +2 \com{Y_{\bar{3}}}{Y_{\bar{4}}}.
\eea
The remaining equation $\Omega \haken \cf=0$ requires
\bea
\label{eq:flow2}
r \lb \dot{Y}_{{4}}-\dot{Y}_{\bar{4}}\rb &=& 2 \com{Y_{1}}{Y_{\bar{1}}} +2 \com{Y_2}{Y_{\bar{2}}} +2 \com{Y_3}{Y_{\bar{3}}} +2 \com{Y_4}{Y_{\bar{4}}}
         - 6 \lb Y_4 -Y_{\bar{4}}\rb.
\eea

{\bf Constant endomorphisms:} For the special case of constant matrices $X_a$, the situation  
corresponds to that of the underlying Sasaki-Einstein manifold $X_{1,1}$ 
with the parameter $\tau$ (or $r$, respectively) just as a label of the foliation along the preferred 
direction of the cone. Gauging the field $X_{\tau}$ to zero, one recovers then from (\ref{eq:flow1}) and (\ref{eq:flow2})
exactly  the additional equivariance conditions
(\ref{eq:eq_rules_q3}), which appeared in the discussion of the gauge theory on $Q_3$. Thus the equivariant gauge theory on $Q_3$ can be considered
as a special instanton solution\footnote{The vanishing of the contributions stemming from the form $e^7$ is  obvious from the Yang-Mills 
action (\ref{eq:YM_action}) and the 
instanton condition (\ref{eq:inst}). Due to $\ast_7 Q \propto e^7$ those terms do not contribute to the action for instanton solutions,
and this is equivalent to the further equivariance conditions (\ref{eq_:eq_q3}).} of the more general setup on $C\lb X_{1,1}\rb$.
%=========================================================
%=========================================================
%=========================================================
%=========================================================
%=========================================================
%=========================================================

\subsection{Moduli space of  $\mathrm{SU}(3)$-equivariant instantons}
\label{sec:moduli_space}
For a desription of the moduli space of the equations (\ref{eq:flow2}) and (\ref{eq:flow1}), (\ref{eq:constraints}) under the equivariance conditions
(\ref{eq:equivariance}), it is advantageous to re-write them in a form similar to the Nahm equations. Then one can employ the techniques used by Donaldson 
\cite{Donaldson84} and Kronheimer \cite{Kronheimer90} for the discussion thereof. We will briefly sketch the application of these methods to 
our system of flow equations, following  \cite{MS15}, where 
the framed moduli space of solutions to the  Hermitian Yang-Mills equations on metric cones over  generic Sasaki-Einstein manifolds is discussed in this way.
Note that the treatment \cite{MS15} uses the canonical connection of \cite{noelle12} as starting point $\Gamma$ for the gauge connection
and that our connection $\Gamma=I_8 \otimes e^8$ in (\ref{eq:ansatz_cone}) differs from it (see Appendix \ref{sec:connections}). 
This is why some modifications, in comparison with \cite{MS15}, will appear in our discussion\footnote{Of course, using the canonical 
connection of \cite{noelle12} yields the results of \cite{MS15} also for $X_{1,1}$. However, for the discussion of the quiver diagrams in Section \ref{sec:qgt}
the connection $\Gamma=I_8 \otimes e^8$ was more suitable because it is valued in the subalgebra $\mathfrak{h}$ and, thus,  adapted 
to the setup of a homogeneous space. The canonical connection, in contrast, is adapted to the Sasaki-Einstein structure of $X_{1,1}$; see Appendix 
\ref{sec:connections}.}.

Changing the argument in the flow equations to 
$\tau =\mathrm{ln}(r)$ and setting\footnote{For the canonical connection (\ref{eq:def_canonical}) of a seven-dimensional 
Sasaki-Einstein manifold, the matrices scale as \cite{MS15}
\bea
Y_{\bar{\a}}= \e^{-\frac{4}{3} \tau} W_{\a}\ \ \ \text{for}\ \ \a=1,2,3\ \ \ \text{and}\  \ \ Y_{\bar{4}} = \e^{-6\tau} Z.
\eea}
\bea
\label{eq:scaling}
Y_{\bar{\a}} \eqqcolon \e^{-\tau} W_{\a}, \ \ \text{for} \ \ \a=1,2, \ \ \ \ 
Y_{\bar{3}} \eqqcolon \e^{-2\tau} W_{3}, \ \ \ \text{and} \ \ \ \ Y_{\bar{4}} \eqqcolon \e^{-6 \tau} Z
\eea
eliminates the linear terms in (\ref{eq:flow1}) and (\ref{eq:flow2}). Defining $s\coloneqq -\frac{1}{6} \e^{-6\tau}=-\frac{1}{6}r^{-6} \in (-\infty,0]$ 
yields Nahm-type equations
\bea
\label{eq:nahm1}
 \frac{\diff W_1}{\diff s} \= 2 \com{W_1}{Z}, \ \ \ \ \ \frac{\diff W_2}{\diff s}\= 2 \com{W_2}{Z},\ \ \ \ \ \frac{\diff W_3}{\diff s}\= 2 \com{W_3}{Z},\\
\label{eq:nahm1b}
 \com{W_1}{W_2} \= 2 W_3 \ \ \ \ \ \ \text{and} \ \ \ \com{W_1}{W_3} \= 0 \= \com{W_2}{W_3},
\eea
(from $\cf^{(2,0)}=0$) and
\bea
\label{eq:nahm2}
&& \mu \lb W_{\a}, Z\rb \coloneqq \frac{\diff}{\diff s} \lb Z+Z^{\+}\rb  + 2 \sum_{\a=1}^3\l^{\a}\lb s\rb \com{W_{\a}}{W_{\a}^{\+}} 
+ 2 \com{Z}{Z^{\+}}\=0
\eea
\normalsize
(from $\Omega \haken \cf=0$), with the non-negative functions
\bea
\label{eq:functions}
\l^1 \lb s\rb \= \l^2 \lb s\rb \coloneqq \lb -6 s\rb^{-\frac{5}{3}} \ \ \ \text{and} \ \ \ \  \l^3 \lb s\rb \coloneqq \lb -6 s\rb^{-\frac{4}{3}}.
\eea
The equation (\ref{eq:nahm2}) shall be referred to as the \emph{real equation} and the equations (\ref{eq:nahm1}) and (\ref{eq:nahm1b}) as \emph{complex equations}. 
The discussion of the moduli space is based
on the invariance of the complex equations under the complexified gauge 
transformation \cite{Donaldson84}
\bea
\label{eq:gauge_transform}
W_{\a} \longmapsto W_{\a}^g \coloneqq g W_{\a} g^{-1}, \ \ \ \text{for}\ \a=1,2,3 \ \ \text{and} \ \ Z \longmapsto Z^g \coloneqq g Z g^{-1} - 
\frac{1}{2} \lb \frac{\diff g}{\diff s}\rb g^{-1}
\eea
with $g \in \mathcal{C} \lb (-\infty, 0], \mathrm{GL}(\C,k)\rb$. A local solution of (\ref{eq:nahm1}) can be attained by applying the gauge
\bea
\label{eq:gauge1}
Z^g \=0 \ \ \ \ \ \Rightarrow \ \ \ \ Z = \frac{1}{2} g^{-1} \frac{\diff g}{\diff s},
\eea
so that -- due to the complex equations (\ref{eq:nahm1}) -- the gauge transformed matrices $W_{\a}^g$ must be constant,
\bea
\label{eq:adj-orbit}
W_{\a} \= g^{-1} T_{\a} g.
\eea
To obtain solutions, one has to choose these constant matrices such that they satisfy (\ref{eq:nahm1b}). One special choice, for instance, 
could be to set $T_3=0$ and take for $T_1$ and $T_2$ elements of a Cartan subalgebra. Note that not only the scaling in (\ref{eq:scaling}) is different
from that in \cite{MS15}, but also the conditions (\ref{eq:nahm1b}): There all three matrices have to commute with each other and, thus, also $T_3$ can be chosen 
as arbitrary element of a Cartan subalgebra. 
 Adapting Donaldson's arguments \cite{Donaldson84, MS15}, 
the real equation (\ref{eq:nahm2}) can be  -- locally on an interval $\mathcal{I} \subset 
(-\infty,0]$ --  considered as the equation of motion (i.e. $\delta \mathcal{L} \propto \m$) of
the Lagrangian
\bea
\label{eq:lagr_real_eq}
\mathcal{L}\= \frac{1}{2} \int_{\mathcal{I}} \diff s \left\{ 2 |Z+Z^{\+}|^2 + 2\l^1(s) |W_1|^2 + 2 \l^2 (s) |W_2|^2  + 2 \l^3 (s)|W_3|^2 \right\}.
\eea
Employing (\ref{eq:gauge1}) and (\ref{eq:adj-orbit}), one can re-write this Lagrangian as \cite{Donaldson84,MS15}
\bea
\mathcal{L}(h) \= \frac{1}{2} \int_{\mathcal{I}} \diff s \left\{\frac{1}{4} \mathrm{tr} \ \lb h^{-1}\frac{\diff h}{\diff s}\rb^2 
 + 2 \sum_{\a=1}^3\l^{\a} \mathrm{tr}\ \lb h T_{\a} h^{-1} T_{\a}^{\+}\rb \right\}  
\ \ \ \ \ \text{with} \ \ \ \ \ \
h \coloneqq g^{\+}g.
\eea
Since the potential term in this Lagrangian is non-negative, the existence of a solution to (\ref{eq:nahm2}) as equation of motion follows 
from a variational problem \cite{Donaldson84}. One still has to ensure some technical aspects: the uniqueness of the solutions, the existence of the gauge transformation 
and the Lagrangian on the \emph{entire} 
interval $(-\infty, 0]$, as well as
the boundedness of $\mu$. In the reference \cite{MS15} these properties are proven, given that for framed instantons, i.e. those with $h=1$ at the 
boundary of the interval $(-\infty,0]$, the following  condition 
\bea
\exists g_0 \in \mathrm{U}(k): \ \lim_{s \rightarrow -\infty} W_{\a} \= \mathrm{Ad}(g_0) T_{\a} 
\eea
is satisfied for constant matrices obeying the conditions (\ref{eq:nahm1b}). For \emph{their} constraints, i.e. mutually commuting matrices $T_{\a}$, it 
is shown that the moduli space can be expressed as diagonal orbit in a product of  coadjoint orbits \cite{MS15}. In our case, however, due to the different constraints (\ref{eq:constraints}), 
the situation might be more involved. But we can at least conclude that (\ref{eq:adj-orbit}) provides local solutions of the Nahm-type \mbox{equations
(\ref{eq:nahm1})-(\ref{eq:nahm2})}.

Moreover, it was shown in the references (see again \cite{MS15})  that the real equation (\ref{eq:nahm2}) can be  considered as a 
moment map $\m: \mathbb{A}^{1,1} \rightarrow \mathrm{Lie}\lb \mathcal{G}_0\rb$ from the space $\mathbb{A}^{1,1}$ of framed solutions to the complex 
equations into the Lie algebra of the framed gauge group $\mathcal{G}_0$. This result still holds here, despite the difference in the connections that 
are used. Hence the moduli space of equivariant Hermitian Yang-Mills instantons on metric cones over 
Sasaki-Einstein manifolds admits the description as K\"ahler quotient \cite{MS15}
\bea
\mathcal{M}= \m^{-1}\lb 0\rb/ {\mathcal{G}_0}.
\eea
%=========================================================
%=========================================================
%=========================================================
%=========================================================
%=========================================================
%=========================================================
\section{Summary and conclusions}

In this article we studied the $\mathrm{SU}(3)$-equivariant dimensional reduction of gauge theories over the Sasaki-Einstein manifold $X_{1,1}$. 
We interpreted the condition of equivariance, which had already occurred in articles \cite{AW11,Haupt15} on $\mathrm{Spin}(7)$-instantons on cones over 
Aloff-Wallach spaces $X_{k,l}$, in terms of quiver diagrams, and we discussed the general construction of the quiver bundles. This yielded
a new class of Sasakian quiver gauge theories. The associated quiver diagram 
of this gauge theory is a \emph{``doubled modified holomorphic chain''}, consisting of two arrows between adjacent vertices and two loops at each vertex, 
and three explicit examples thereof were considered in the article.  For the comparison with the  gauge theory on the underlying K\"ahler manifold $Q_3$
 we studied 
the quivers also in the entire weight diagram of $G=\mathrm{SU}(3)$, which implied degeneracies of the arrows. This behavior is 
similar to the case  \cite{LPS06,T11} of the five-dimensional Sasaki-Einstein manifold $T^{1,1}$ over $\C P^1 \times \C P^1$. The reduction to the gauge theory 
on $Q_3$ led to the correct, expected result for the quiver diagram \cite{SU3}: the weight diagram of $\mathrm{SU}(3)$.
 
For the investigation of the vacua described by this gauge theory we imposed the Hermitian Yang-Mills equations
on the metric cone $C\lb X_{1,1}\rb$. The resulting flow equations have been  re-written in a form similar to Nahm's equations, which 
allowed a discussion based on Kronheimer's \cite{Kronheimer90} and Donaldson's  \cite{Donaldson84} work and its generalized application 
to equivariant HYM instantons on Calabi-Yau cones \cite{MS15}.
Since we formulated the quiver gauge theory by using an instanton connection different from that of \cite{noelle12} in the gauge connection,
some modifications appeared. While the real equation can be still interpreted as a moment map for framed instanton solutions, as in \cite{MS15}, 
and, thus, leads to a description of the moduli space as a K\"ahler quotient, the description based on coadjoint orbits is more involved: The HYM equations 
impose a non-trivial commutation relation on the gauge transformed matrices, in contrast to \cite{MS15}, where they have to commute with each other. Thus,
the behavior is more complicated and further effort would be needed to study the consequences thereof in detail.

%==========================================================
%==========================================================
%==========================================================
%==========================================================
%==========================================================
%==========================================================
\section*{Acknowledgements}
The author thanks Olaf Lechtenfeld, Alexander Popov, and Marcus Sperling for fruitful discussions and comments. This work was  
done within the project supported by the  Deutsche Forschungsgemeinschaft (DFG, Germany) under the grant LE 838/13 and 
was supported by the Research Training Group RTG 1463 ``Analysis, Geometry and String Theory'' (DFG).

%==========================================================
%==========================================================
%==========================================================
%==========================================================
%==========================================================
%==========================================================
\appendix
\section{Appendix}
\subsection{$\mathrm{SU}(3)$ generators and structure constants}
\label{sec:geometry_aw}
The generators defined by the choice of the \mbox{1-forms} in (\ref{eq:def_1forms}) read 
\small
\bea
\label{eq:generators}
 &&I_1^- \coloneqq \sqrt{2} \begin{pmatrix}
                                    0 & 0 &0\\
                                    0 & 0 &0\\
                                    1 & 0 &0
                                   \end{pmatrix}, \ \ \
 I_2^- \coloneqq \sqrt{2} \begin{pmatrix}
                                    0 & 1 &0\\
                                    0 & 0 &0\\
                                    0 & 0 &0
                                   \end{pmatrix}, \ \ \
 I_3^- \coloneqq  \begin{pmatrix}
                                    0 & 0 &0\\
                                    0 & 0 &0\\
                                    0 & 1 &0
                                   \end{pmatrix}, \ \ \
 I_7 \coloneqq \im  \begin{pmatrix}
                                    0 & 0 &0\\
                                    0 & -1 &0\\
                                    0 & 0 &1
                                   \end{pmatrix},\\
\nonumber &&I_{\bar{1}}^+ \coloneqq \sqrt{2} \begin{pmatrix}
                                    0 & 0 &-1\\
                                    0 & 0 &0\\
                                    0 & 0 &0
                                   \end{pmatrix}, \ \ 
 I_{\bar{2}}^+ \coloneqq \sqrt{2} \begin{pmatrix}
                                    0 & 0 &0\\
                                    -1 & 0 &0\\
                                    0 & 0 &0
                                   \end{pmatrix}, \ \ \
 I_{\bar{3}}^+ \coloneqq  \begin{pmatrix}
                                    0 & 0 &0\\
                                    0 & 0 &-1\\
                                    0 & 0 &0
                                   \end{pmatrix}, \ \ \
 I_8 \coloneqq \frac{\im}{\sqrt{3}}  \begin{pmatrix}
                                    2 & 0 &0\\
                                    0 & -1 &0\\
                                    0 & 0 &-1
                                   \end{pmatrix},
\eea
\normalsize
and we define the  structure constants via the commutation relations
\bea
\nonumber \com{-\im I_j}{I^-_{\a}} &=& C_{j\a}^{\b} I_{\b}^-, \ \ \
\com{-\im I_j}{I_{\bar{\a}}^+} \= C_{j\bar{\a}}^{\bar{\b}} I_{\bar{\b}}^+, \ \ \
\com{I_{\a}^-}{I_{\b}^-} \= C_{\a \b}^{\g} I_{\g}^-,\\
\com{I_{\bar{\a}}^+}{I_{\bar{\b}}^+}&=& C_{\bar{\a}\bar{\b}}^{\bar{\g}}I_{\bar{\g}}^+, \ \ \
\com{I_{\a}^-}{I_{\bar{\b}}^+}\= - \im C_{\a \bar{\b}}^j I_j + C_{\a \bar{\b}}^{\g} I_{\g}^- + C_{\a \bar{\b}}^{\bar{\g}}I_{\bar{\g}}^+.
\eea
The non-vanishing structure constants are  \cite{AW11}
\begin{align}
\label{eq:struc_const_aw}
\nonumber C_{3\bar{2}}^1 &= -C_{3\bar{1}}^{2} \= -1 \= -C_{2\bar{3}}^{\bar{1}}\= C_{1\bar{3}}^{\bar{2}},  & 
          C_{12}^{3}& = 2 \= C_{\bar{1}\bar{2}}^{\bar{3}}, \\
          C_{71}^1 &=  C_{72}^2 \= 1\=  - C_{7\bar{1}}^{\bar{1}}\= -C_{7\bar{2}}^{\bar{2}}, &
          C_{73}^3 &= 2 \= -C_{7\bar{3}}^{\bar{3}}, \\
\nonumber C_{81}^1 &= - C_{82}^2\= -\sqrt{3} \= - C_{8\bar{1}}^{\bar{1}} \=C_{8\bar{2}}^{\bar{2}} , &
          C_{83}^3 &= 0 \= C_{8\bar{3}}^{\bar{3}}, \\
\nonumber C_{1\bar{1}}^7 &=
          C_{2\bar{2}}^7 \= 
          C_{3\bar{3}}^7 \= -1, &
          C_{1\bar{1}}^8 &= -C_{2\bar{2}}^8  \= \sqrt{3}.
\end{align}

By the Maurer-Cartan equations,
\bea
\diff \Theta^{\a} &=& - \im C_{j\b}^{\a}- \frac{1}{2} C_{\b \g}^{\a} \Theta^{\b\g}- C_{\b\bar{\g}}^{\a} \Theta^{\b\bar{\g}}, \ \ \ \ \ \ \ \
\diff e^j \= \im C_{\b\bar{\g}}^{j} \Theta^{\b\bar{\g}},
\eea
 they yield again the structure equations (\ref{eq:structure_eqs_aw}). In terms of real forms
\bea
\Theta^1 \eqqcolon e^1 - \im e^2, \ \ \ \Theta^2 \eqqcolon e^3 - \im e^4, \ \ \text{and}\ \ \  \Theta^3 \eqqcolon e^5 - \im e^6,
\eea
the structure equations read
\begin{align}
\nonumber \diff e^1 &= \sqrt{3}e^{82}- e^{72}- e^{35}-e^{46},& \diff e^2 &= -\sqrt{3}e^{81}+e^{71}-e^{36}+e^{45},\\
\nonumber \diff e^3 &= -\sqrt{3}e^{84}-e^{74}+e^{15}+e^{26},& \diff e^4 &= \sqrt{3}e^{83} +e^{73}+ e^{16}-e^{25},\\ 
\nonumber \diff e^5 &= -2 e^{76}-2e^{13}-2e^{24}, & \diff e^6 &= 2 e^{75} -2e^{14} - 2e^{23},\\
          \diff e^7 &= 2 e^{12} + 2^{34} + 2^{56}, & \diff e^8 &= -2 \sqrt{3}e^{12} + 2\sqrt{3} e^{34}.
\end{align}
%==========================================================
%==========================================================
%==========================================================
%==========================================================
%==========================================================
%==========================================================
\subsection{Connections and instanton equation}
\label{sec:connections}

On the homogeneous space $X_{1,1}=G/H=\mathrm{SU}(3)/\mathrm{U}(1)_{1,1}$ we consider the connection with torsion
\bea
\label{eq:tor_can}
T \lb X,Y \rb \coloneqq - \com{X}{Y}_{\mathfrak{m}}
\eea
for vector fields $X$, $Y$ on $G/H$, where $\com{\cdot}{\cdot}_{\mathfrak{m}}$ denotes the projection of the commutator to the complement $\mathfrak{m}$; 
this yields the  following torsion components
\bea
T^{\m}_{\rho \sigma} \= -C_{\rho \sigma}^{\m} \ \ \ \ \ \ \ \text{for} \ \ \mu, \rho, \sigma \= 1, \ldots,7.
\eea
Using the structure equations and the Maurer-Cartan equation
\bea
\diff e^{\m} &=& -\frac{1}{2} C_{\rho \sigma}^{\m} e^{\rho \sigma} \= - C_{8\rho}^{\m} e^{8} \wedge e^{\rho} +\frac{1}{2} T^{\mu}_{\rho \sigma} e^{\rho \sigma}\\
     \nonumber &\eqqcolon & - \Gamma^{\m}_{\rho} \wedge e^{\rho} +T^{\mu},
\eea
one obtains the conection \mbox{1-forms}
\bea
\Gamma^{\m}_{\rho} \= C_{8 \rho}^{\m} e^8 \ \ \ \ \ \ \Rightarrow \Gamma \= I_8 \otimes e^8,
\eea
which is  the  $\mathrm{U}(1)$-connection used in the ansatz for the gauge connection in (\ref{eq:ansatz_cone}). Its curvature
\bea
\cf_{\Gamma} = \diff \Gamma +\Gamma \wedge \Gamma = 2 \sqrt{3} I_8\lb e^{12}-e^{34}\rb 
\eea
satifies the instanton equation
\bea
\ast_7 \cf_{\Gamma} \= - \lb e^{12}+e^{34}+e^{56} \rb \wedge e^7 \wedge \cf_{\Gamma} \= - \ast_7 Q \wedge \cf_{\Gamma}.
\eea
for the \mbox{4-form} $Q=e^{1234}+e^{1256}+e^{3456}$ from (\ref{eq:Q_SE}). Because of
\bea
\lb \diff \ast_7 Q \rb \wedge \cf_{\Gamma} \propto  \lb e^{1234}+e^{1256}+e^{3456}\rb \wedge \lb e^{12}-e^{34}\rb =0
\eea
the torsion term in  (\ref{eq:YMT}) vanishes, so that the usual torsion-free Yang-Mills equation is obtained. 
This is the intention of using special geometric structures.
Note, however, that our $\mathrm{U}(1)$-connection does not coincide with 
what is defined as \emph{canonical connection of a Sasaki-Einstein manifold} in \cite{noelle12}. Its torsion for a seven-dimensional 
Sasaki-Einstein manifold is defined via
\bea
\label{eq:def_canonical}
T^a = \frac{2}{3} P_{a\m \n} e^{\m \n}\ \ \text{for} \ \ a=1, \ldots, 6  \ \ \ \ \ \text{and}  \ \ \ \  \ \ T^7 = P_{7\m\n} e^{\m\n} \ \ \
\text{with} \ \ \
P \coloneqq \eta \wedge \omega = e^7 \wedge \omega.
\eea
Since this definition does not require a homogeneous space, but only exploits the Sasaki-Einstein structure,
it  allows for general discussions of gauge theories on those spaces, as used for example in \cite{MS15,IP12}. On $X_{1,1}$ this 
canonical connection is expressed by the connection matrix
\bea
\diff \begin{pmatrix} 
       \Theta^1\\
       \Theta^2\\
       \Theta^3\\
       e^7
      \end{pmatrix}
\=
\begin{pmatrix}
 \frac{1}{3} \im e^7 + \sqrt{3} \im e^8 	&0	&-\Theta^{\bar{2}}	&0\\
0						&\frac{1}{3} \im e^7 - \sqrt{3}\im e^8 & \Theta^{\bar{1}} & 0\\
\Theta^2    & -\Theta^1 & -\frac{2}{3}\im e^7 & 0\\
0 & 0 &0 &0
\end{pmatrix}
\wedge 
\begin{pmatrix} 
       \Theta^1\\
       \Theta^2\\
       \Theta^3\\
       e^7
      \end{pmatrix}+
\vec{T}.
\eea
Thus, the canonical connection is adapted to the $\mathrm{SU}(3)$ structure of $X_{1,1}$.
 
On the metric cone (with the rescaled forms  $\tilde{e}^{\m} \coloneqq r e^{\m}$) or on the 
conformally equivalent cylinder, respectively, the connection $\Gamma=I_8 \otimes e^8$ is still an instanton for the form 
\bea
Q_Z \= \frac{1}{2} \Omega \wedge \Omega &=& r^4 \lb e^{1234}+e^{1256}+e^{12\tau 7}+e^{3456}+e^{34\tau 7}+ e^{56\tau 7}\rb \\
  \nonumber                                       &= &  \tilde{e}^{1234}+\tilde{e}^{1256}+\tilde{e}^{12\tau 7}+\tilde{e}^{3456}+\tilde{e}^{34\tau 7}
                                                   + \tilde{e}^{56\tau 7} \= \ast_8 Q_Z
\eea
because we have
\bea
\ast_8 \lb \tilde{e}^{12}- \tilde{e}^{34}\rb \= - \lb \tilde{e}^{12}-\tilde{e}^{34}\rb \wedge \tilde{e}^{56\tau 7} \= - Q_Z \wedge \lb \tilde{e}^{12}-
\tilde{e}^{34}\rb.
\eea
%==========================================================
%==========================================================
%==========================================================
%==========================================================
%==========================================================
%==========================================================
\subsection{Details of the moduli space description}

This section  provides some technical aspects of the description in Section \ref{sec:moduli_space}. For details, the reader should consult the references, 
in particular \cite{MS15}.
To show that the real equation follows (over some range) as equation of motion of the Lagrangian (\ref{eq:lagr_real_eq}), one considers \cite{Donaldson84} 
the variation of 
the matrices $W_a$ with respect to $g$ close to the identity. Writing $g=1+ \delta g$, where $\delta g$ is self-adjoint, one obtains from the 
gauge transformation (\ref{eq:gauge_transform})
\bea
\label{eq:var1}
\delta W_{\a} &=& \lb 1+ \delta g\rb W_{\a} \lb 1+\delta g\rb^{-1} -W_{\a} \= \com{\delta g}{W_{\a}} \ \ \ \ \text{for} \ \ \ \ \a=1,2,3
\eea 
and 
\bea
\label{eq:var2}
 \delta Z \= \lb 1+\delta g \rb Z \lb 1+\delta g\rb^{-1} -Z -\frac{1}{2} \frac{\diff}{\diff s} \lb 1+\delta g\rb \lb 1+\delta g\rb^{-1}
        \= \com{\delta g}{Z} -\frac{1}{2} \frac{\diff}{\diff s} \delta g.
\eea
Using the result (\ref{eq:var1}), one derives the following variation 
\bea
\label{eq:var3}
\nonumber \delta \int \diff s\ |W_{\a}|^2 &\coloneqq& \delta \int \diff s\ \mathrm{tr}\ W_{\a} W_{\a}^{\+} \= 
2 \mathrm{Re} \int \diff s\ \tr\ \delta \lb W_{\a}\rb  W_{\a}^{\+}
\= 2 \mathrm{Re} \int \diff s\ \tr\ \com{\delta g}{W_{\a}} W_{\a}^{\+}\\
 &=& 2 \mathrm{Re} \int \diff s\ \mathrm{tr}\ \delta g \com{W_{\a}}{W_{\a}^{\+}} \ \ \ \ \ \text{for} \ \ \ \a=1,2,3
\eea
and 
\bea
\label{eq:var4}
\nonumber \delta \int \diff s\ |Z+Z^{\+}|^2 &=& 2\mathrm{Re} \int \diff s\ \mathrm{tr}\ \lb \com{\delta g}{Z- Z^{\+}} - \frac{\diff}{\diff s} \delta g 
\rb \lb Z +Z^{\+}\rb \\
&=& 2 \mathrm{Re} \int \diff s\ \mathrm{tr} \delta g \lb \frac{\diff}{\diff s} \lb Z+Z^{\+}\rb + 2 \com{Z}{Z^{\+}}\rb.
\eea
Putting the results from (\ref{eq:var3}) and (\ref{eq:var4}) together with the prefactors $\l^{\a}(s)$, (\ref{eq:functions}) yields the 
Lagrangian (\ref{eq:lagr_real_eq}) and
shows that the real equation is the equation of motion thereof. That the Lagrangian can be defined for the entire range $s \in (-\infty,0]$ and other 
technical issues
can be found in \cite{MS15}. The only quantitive difference is the concrete form of the factors $\l^{\a}(s)$ but this does not affect the general line of
reasoning.

%==========================================================
%==========================================================
%==========================================================
%==========================================================
%==========================================================
%==========================================================
%==========================================================
%==========================================================
%==========================================================


\begin{thebibliography}{200}
%\addtolength{\itemsep}{-1pt}
{\footnotesize
\bibitem{AHS78}
M.F.~Atiyah, N.J.~Hitchin, and I.M.~Singer,
``Self-duality in four-dimensional Riemannian geometry'',
Proc. R. Soc. Lond. A. {\bf 362} (1978) 425.

\bibitem{Corrigan83}
E.~Corrigan, C.~Devchand, D.B.~Fairlie, and J.~Nuyts,
``First-order equations for gauge fields in spaces of dimension 
greater than four'',
Nucl. Phys. B {\bf 214} (1983) 452.

\bibitem{Ward84}
R.S.~Ward,
``Completely solvable gauge-field equations in dimensions greater than four'',
Nucl. Phys. B {\bf 236} (1984) 381.

\bibitem{Hull}
C.M. Hull,
"Higher dimensional Yang-Mills
theories and topological terms",
Adv. Theor. Math. Phys. {\bf 2} (1998) 619,
[arXiv:hep-th/9710165v2].
 
\bibitem{noelle12}
D.~Harland and C.~N\"olle,
``Instantons and Killing spinors'',
JHEP {\bf 03} (2012) 082,
[arXiv:1109.3552].

\bibitem{heterotic}
U.~Gran, G.~Papadopoulos, and D.~Roest,
``Supersymmetric heterotic string backgrounds'',
Phys. Lett. B {\bf 656} (2007) 119,
[arXiv:0706.4407].

\bibitem{Coset92}
D.~Kapetanakis and G.~Zoupanos,
``Coset space dimensional reduction of gauge theories'',
Phys. Rept. {\bf 219} (1992) 4.

\bibitem{GP02}
L.~Alvarez-C\'onsul and O.~Garc\'ia-Prada,
``Dimensional reduction and quiver bundles'',
J. Reine Angew. Math. {\bf 556} (2003) 1,
[arXiv:math-dg/0112160].

\bibitem{SL2C}
L.~Alvarez-C\'onsul and O.~Garc\'ia-Prada,
``Dimensional reduction, $\mathrm{SL}(2,\C)$-equivariant bundles 
and stable holomorphic chains'',
Int. J. Math. {\bf 12} (2001) 159,
[arXiv:math-dg/0112159].


\bibitem{DS11}
B.P.~Dolan and R.J.~Szabo,
``Equivariant dimensional reduction and quiver gauge theories'',
 Gen. Rel. Grav. {\bf 43} (2010) 2453,
[arXiv:hep-th/1001.2429].


\bibitem{QGT}
O.~Lechtenfeld, A.D.~Popov, and R.J.~Szabo,
``Quiver gauge theory and noncommutative vortices'',
Prog. Theor. Phys. Suppl. {\bf 171} (2007) 258,
[arXiv:0706.0979]

\bibitem{PS06}
A.D.~Popov and R.J.~Szabo,
``Quiver gauge theory of nonabelian vortices
and noncommutative instantons in higher dimensions'',
J. Math. Phys. {\bf 47} (2006) 012306,
[arXiv:hep-th/0504025].

\bibitem{Bismas}
I.~Biswas,
``Holomorphic Hermitian vector bundles over the Riemann sphere'',
Bull. Sci. Math. {\bf 132} (2008) 246.

\bibitem{Dolan:2009ie}
  B.P.~Dolan and R.J.~Szabo,
  ``Dimensional reduction, monopoles and dynamical symmetry breaking'',
  \mbox{JHEP {\bf 03}} (2009) 059,
  [arXiv:0901.2491].

% \bibitem{Szabo:2014zua}
%   R.J.~Szabo and O.~Valdivia,
%   ``Covariant quiver gauge theories'',
%   JHEP {\bf 06} (2014) 144,
%   [arXiv:1404.4319].



\bibitem{LPS06}
O.~Lechtenfeld, A.D.~Popov, and R.J.~Szabo,
``Rank two quiver gauge theory, graded connections and noncommutative vortices,''
JHEP {\bf 09} (2006) 054,
[arXiv:hep-th/0603232].

\bibitem{SU3}
O.~Lechtenfeld, A.D.~Popov, and R.J.~Szabo,
``$\mathrm{SU}(3)$-equivariant quiver gauge theories and nonabelian vortices'',
JHEP {\bf 08} (2008) 093,
[arXiv:0806.2791v2].

\bibitem{BG}
C.~Boyer and K.~Galicki,
\emph{Sasakian Geometry} (Oxford University Press, 2008).



\bibitem{Sparks}
J. Sparks,
``Sasaki-Einstein manifolds'',
Surv. Diff. Geom. {\bf 16} (2011) 265,
[arXiv:1004.2461]

\bibitem{Joyce}
D.~Joyce,
``Lectures on Calabi-Yau and special Lagrangian geometry'',
Preprint arXiv: math/0108088.
M.~Gross, D.~Huybrechts, and D.~Joyce,
{\sl Calabi-Yau manifolds and related geometries},
(Springer, 2003).

\bibitem{Greene}
B.R.~Greene,
``String theory on Calabi-Yau manifolds'',
Preprint arXiv: hep-th/9702155.

\bibitem{double_quiver}
A.D.~Popov and R.J.~Szabo,
``Double quiver gauge theory and nearly Kahler flux compactifications'',
JHEP {\bf 02} (2012) 033,
[arXiv:1009.3208v2].


\bibitem{Lechtenfeld:2014fza}
  O.~Lechtenfeld, A.D.~Popov, and R.J.~Szabo,
  ``Sasakian quiver gauge theories and instantons on Calabi-Yau cones'',
  Preprint arXiv:1412.4409.

\bibitem{S5}
O.~Lechtenfeld, A.D.~Popov, M.~Sperling, and R.J.~Szabo,
``Sasakian quiver gauge theories and instantons on cones over lens 5-spaces'',
Nucl. Phys. B {\bf 899} (2015) 848,
[arXiv:1506.02786].

\bibitem{T11}
J.C.~Geipel, O.~Lechtenfeld, A.D.~Popov, and R.J.~Szabo,
``Sasakian quiver gauge theories and instantons on the conifold'',
Nucl. Phys. B (2016), doi:10.1016/j.nuclphysb.2016.04.016,
[arXiv:1601.05719].


\bibitem{SaEin_S2S3}
J.P.~Gauntlett, D.~Martelli, J.~Sparks, and D.~Waldram,
``Sasaki-Einstein metrics on $S^2 \times S^3$'',\\
Adv. Theor. Math. Phys. {\bf 8} (2004) 711,
[arXiv:hep-th/0403002].

\bibitem{Martelli:2004}
  D.~Martelli and J.~Sparks,
  ``Toric geometry, Sasaki-Einstein manifolds and a new infinite class of AdS/CFT duals'',
  Commun.\ Math.\ Phys.\  {\bf 262} (2006) 51,
  [arXiv:hep-th/0411238].





\bibitem{AW75}
S.~Aloff and N.R.~Wallach,
``An infinite family of distinct 7-manifolds admitting positively curved
Riemannian structures'',
Bull. Amer. Math. Soc. {\bf 81} (1975) 93.

\bibitem{CMS94}
F.M.~Cabrera, M.D.~Monar, and A.F.~Swann,
``Classification of $G_2$-Structures'',
J. London Math. Soc. (2) {\bf 53} (1996) 407.

\bibitem{FKMS97}
T.~Friedrich, I.~Kath, A~Moroianu, and U.~Semmelmann,
``On nearly parallel $G_2$-structures'',
J. Geom. Phys. {\bf 23} (1997) 259.

\bibitem{AW11}
A.S.~Haupt, T.A.~Ivanova, O. Lechtenfeld, and A.D.~Popov,
``Chern-Simons flows on Aloff-Wallach spaces and 
$\mathrm{Spin}(7)$ instantons'',
Phys. Rev. D {\bf 83} (2011) 105028,
[arXiv:1104.5231v1].


\bibitem{Haupt15}
A.S.~Haupt,
''Yang-Mills solutions and $\mathrm{Spin}(7)$-instantons on cylinders over coset
spaces with $G_2$-structure``, JHEP {\bf 03} (2016) 038, [arXiv:1512.07254v1].

\bibitem{sugra}
L.~Castellani, L.J.~Romans, and N.P.~Warner,
``A classification of compactifying solutions for $d=11$
supergravity'',
Nucl. Phys. B {\bf 241} (1984) 429.

\bibitem{Donaldson84}
S.K.~Donaldson,
``Nahm's equations and the classification of monopoles'',
Commun. Math. Phys. {\bf 96} (1984) 387.

\bibitem{Kronheimer90}
P.B.~Kronheimer,
``A hyper-K\"ahlerian structure on coadjoint orbits of a semisimple complex group'',
J. London Math. Soc. {\bf 42} (1990) 193.




\bibitem{MS15}
M.~Sperling,
``Instantons on Calabi-Yau cones'',
Nucl. Phys. B {\bf 901} (2015) 354,
[arXiv:1505.01755].



\bibitem{CGM}
C.P.~Boyer, K.~Galicki, and B.M.~Mann,
``The geometry and topology of 3-Sasakian manifolds'',
J. reine angew. Math. {\bf 455} (1994) 183.

\bibitem{stringy}
B.~Florea, S.~Kachru, J.~McGreevy, and N.~Saulina,
``Stringy instantons and quiver gauge theories'',
JHEP {\bf 05} (2007), 024,
[arXiv:hep-th/0610003v3].


\bibitem{IP12}
T.A.~Ivanova and A.D.~Popov,
``Instantons on special holonomy manifolds'',
Phys. Rev. D {\bf 85} (2012) 105012,
[arXiv:1203.2657].

\bibitem{conical14}
S.~Bunk, O.~Lechtenfeld, A.D.~Popov, and M.~Sperling,
``Instantons on conical half-flat 6-manifolds'',
\mbox{JHEP {\bf 01}} (2015) 030,
[arXiv:1409.0030].

\bibitem{Bunk14}
S.~Bunk, T.A.~Ivanova, O.~Lechtenfeld, A.D.~Popov, and M.~Sperling,
``Instantons on sine-cones over Sasakian manifolds'',
Phys. Rev. D {\bf 90} (2014) 065028,
[arXiv:1407.2948].

\bibitem{Lubbe}
I.~Bauer, T.A.~Ivanova, O.~Lechtenfeld, and F.~Lubbe,
``Yang-Mills instantons and dyons on homogeneous $G_2$-manifolds'',
JHEP {\bf 10} (2010) 044,
[arXiv:1006.2388].

\bibitem{Popov09}
A.D.~Popov,
``Hermitian Yang-Mills equations and pseudo-holomorphic bundles on nearly 
K\"ahler and nearly Calabi-Yau twistor 6-manifolds'', 
Nucl. Phys.  B {\bf 828} (2010) 594, 
[arXiv:0907.0106].

\bibitem{quiver1}
H.~Derksen and J.~Weyman,
``Quiver representations'',
Notices of the AMS {\bf 52}(2) (2005) 200.

\bibitem{quiver2}
R.~Schiffler,
{\sl Quiver representations}, 
(Springer, 2014).

\bibitem{KN}
S.~Kobayashi and K.~Nomizu,
{\sl Foundations of Differential Geometry}, Volume 1
(Interscience Publishers, 1963).

\bibitem{FH}
W.~Fulton and J.~Harris,
{\sl Representation Theory},
(Springer, 1991).

\bibitem{Donaldson85}
S.K~Donaldson,
``Anti-self dual Yang-Mills connections over complex algebraic surfaces and stable vector bundles'',
Proc. Lond. Math. Soc. {\bf 50} (1985) 1.

\bibitem{UY86}
K.~Uhlenbeck and S.-T. Yau,
``On the existence of Hermitian Yang-Mills connections in stable vector bundles'',
Commun. Pure Appl. Math. {\bf 39} (1986) 257.














% \bibitem{Biquard96}
% O.~Biquard,
% ``Sur les \'equations de Nahm et la structure de Poisson des alg\`ebres de Lie semi-simples complexes'',
% Math. Ann. {\bf 304} (1996) 253.


}
\end{thebibliography}
\end{document}